\documentclass[a4paper,fleqn]{cas-dc}

\usepackage[numbers]{natbib}

\usepackage{algorithmicx}
\usepackage{amsmath,amssymb,amsfonts}
\usepackage{textcomp}
\usepackage{xcolor}
\usepackage{algorithm}
\usepackage{algpseudocode}
\usepackage{graphicx}
\usepackage{subcaption}
\usepackage{cleveref}

\def\BibTeX{{\rm B\kern-.05em{\sc i\kern-.025em b}\kern-.08em
    T\kern-.1667em\lower.7ex\hbox{E}\kern-.125emX}}

\graphicspath{{images/}{images/section2-background/}{images/section3-rollback/}{images/section4-unr/}{images/section5-routing/}{images/section6-experiments/}{images/biography/}}

\ExplSyntaxOn
\keys_define:nn { cas / fig }
  {
    t    .code:n = \tl_set:Nn \l_fig_pos_tl {t},
    b    .code:n = \tl_set:Nn \l_fig_pos_tl {b},
    p    .code:n = \tl_set:Nn \l_fig_pos_tl {p},
    h    .code:n = \tl_set:Nn \l_fig_pos_tl {h},
    ht   .code:n = \tl_set:Nn \l_fig_pos_tl {ht},
    hb   .code:n = \tl_set:Nn \l_fig_pos_tl {hb},
    tp   .code:n = \tl_set:Nn \l_fig_pos_tl {tp},
    tb   .code:n = \tl_set:Nn \l_fig_pos_tl {tb},
    htb  .code:n = \tl_set:Nn \l_fig_pos_tl {htb},
    tbp  .code:n = \tl_set:Nn \l_fig_pos_tl {tbp},
    htbp .code:n = \tl_set:Nn \l_fig_pos_tl {htbp}
  }
\keys_define:nn { cas / tbl }
  {
    t    .code:n = \tl_set:Nn \l_tbl_pos_tl {t},
    b    .code:n = \tl_set:Nn \l_tbl_pos_tl {b},
    p    .code:n = \tl_set:Nn \l_tbl_pos_tl {p},
    h    .code:n = \tl_set:Nn \l_tbl_pos_tl {h},
    ht   .code:n = \tl_set:Nn \l_tbl_pos_tl {ht},
    hb   .code:n = \tl_set:Nn \l_tbl_pos_tl {hb},
    tp   .code:n = \tl_set:Nn \l_tbl_pos_tl {tp},
    tb   .code:n = \tl_set:Nn \l_tbl_pos_tl {tb},
    htb  .code:n = \tl_set:Nn \l_tbl_pos_tl {htb},
    tbp  .code:n = \tl_set:Nn \l_tbl_pos_tl {tbp},
    htbp .code:n = \tl_set:Nn \l_tbl_pos_tl {htbp}
  }
\ExplSyntaxOff

\begin{document}
\let\WriteBookmarks\relax
\def\floatpagepagefraction{1}
\def\textpagefraction{.001}

\shorttitle{OACM: Optimistic Asynchronous Communication Model for Large-Scale SNN Simulation}

\shortauthors{Feng et~al.}

\title [mode = title]{OACM: Optimistic Asynchronous Communication Model for Large-Scale SNN Simulation}

\author[1]{Guangnan Feng}[orcid=0000-0002-1382-280X]

\ead{fenggn7@mail.sysu.edu.cn}

\credit{Conceptualization, Methodology, Investigation, Writing – review \& editing, Supervision}

\author[1]{Yangle Zeng}[orcid=0009-0003-2512-9839]

\ead{zengyle@mail2.sysu.edu.cn}

\credit{Methodology, Software, Writing – original draft}

\author[1]{Honghui Liang}[orcid=0009-0006-6509-3000]

\ead{lianghh33@mail2.sysu.edu.cn}

\credit{Software, Investigation}

\author[1]{Zhiguang Chen}[orcid=0000-0002-9318-5715]

\ead{chenzhg29@mail.sysu.edu.cn}

\credit{Project administration}

\author[1]{Yutong Lu}[orcid=0000-0001-5315-3375]

\cormark[1]

\ead{luyutong@mail.sysu.edu.cn}

\credit{Funding acquisition}

\affiliation[1]{organization={Sun Yat-sen University},
            city={Guangzhou},
            postcode={510006},
            country={China}}

\cortext[1]{Corresponding author}

\begin{abstract}
Spiking Neural Network (SNN) simulation serves as a crucial tool for understanding brain dynamics and advancing neuromorphic computing, but faces significant scalability challenges in large-scale distributed implementations.
The primary bottlenecks arise from frequent synchronization overhead in time-driven simulators and extensive secondary rollbacks in optimistic PDES approaches, compounded by inefficient communication patterns that underutilize network bandwidth.

In this paper, we propose OACM, an Optimistic Asynchronous Communication Model that addresses these challenges through three key innovations.
First, we design Optimistic Hybrid SNN Simulation that combines the implementation simplicity of time-driven approaches with reduced synchronization frequency through strategic rollback mechanisms within synchronization windows.
Second, we implement asynchronous one-sided communication using the UNR library, eliminating handshake latency and achieving complete computation-communication overlap.
Third, we develop adaptive message aggregation and routing strategies within a 2D-HyperX virtual topology to optimize bandwidth utilization for small message traffic.

Experimental evaluation on a high-performance computing cluster demonstrates that OACM achieves up to 1.4$\times$ speedup over the original CORTEX simulator and over 22.6$\times$ speedup compared to NEST when simulating a multi-area marmoset brain model at scales of up to 174 compute nodes.
\end{abstract}


\maketitle

\section{Introduction}

Spiking Neural Network (SNN) simulation is pivotal for bridging computational neuroscience and artificial intelligence, offering a biologically realistic model to decipher brain dynamics and inspire machine intelligence.
The brain's unparalleled energy efficiency and robust learning capabilities provide a crucial blueprint for overcoming key bottlenecks in modern AI, such as its massive data dependency and lack of generalization, which hinder progress toward Artificial General Intelligence.
In contrast, the human brain operates with remarkable energy efficiency (approximately 20 watts), seamlessly integrates multimodal information, and exhibits robust generalization. 
Additionally, understanding and treating neurological disorders such as Alzheimer's and Parkinson's diseases necessitates a system-level investigation of aberrant neural activity, a pursuit for which large-scale SNN simulation is an indispensable tool.

As researchers strive to emulate SNNs at the scale of mammalian brains, which encompass 10 million to 100 billion neurons interconnected through trillions of synapses \cite{igarashi2019large, yamaura2020simulation, du2022low, lu2023digital}, the computational complexity grows exponentially.
This scaling challenge is further compounded by biologically plausible models requiring millisecond-level temporal resolution and intricate synaptic dynamics.
Consequently, the development of high-performance SNN simulators capable of balancing biological fidelity with computational efficiency has emerged as a critical frontier in computational neuroscience and neuromorphic engineering.

To meet these demands, modern simulators increasingly adopt parallel acceleration paradigms, achieving remarkable progress in scalability and performance.
Time-driven simulators like NEST \cite{Gewaltig:NEST}, CORTEX \cite{lyu2024cortex}, and MONET \cite{igarashi2019large} have demonstrated exceptional scalability through their straightforward implementation and efficient parallelization.
Event-driven approaches and optimistic PDES techniques have shown promise in reducing computational overhead by updating only upon spike events and relaxing synchronization constraints.
Communication optimizations, including asynchronous protocols \cite{igarashi2019large, fernandez2019communication}, dedicated communication threads \cite{lyu2024cortex}, and virtual topology routing \cite{liu2022regularizing, du2024hrcm}, have significantly improved computation-communication overlap and bandwidth utilization.
These advancements collectively enable simulations at unprecedented scales, bringing the field ever closer to whole-brain emulation.

Despite these significant achievements, fundamental challenges persist in synchronization and communication efficiency.
Time-driven simulators suffer from the overhead of frequent global synchronization, while optimistic PDES approaches face extensive secondary rollbacks under load imbalance.
Existing communication implementations, whether synchronous or asynchronous, introduce implicit synchronization overhead through handshake protocols or pre/post-synchronization requirements.
Moreover, the irregular communication patterns in SNN simulations generate numerous small messages that severely underutilize network bandwidth.
These bottlenecks collectively limit the scalability and efficiency of large-scale SNN simulations, demanding more sophisticated approaches to synchronization and communication.

To address these challenges, we propose OACM, an Optimistic Asynchronous Communication Model for large-scale SNN simulation, and make the following contributions:

\begin{itemize}

\item We design Optimistic Hybrid SNN Simulation, avoiding over-synchronization in time-driven simulators and the spread of secondary rollbacks in Optimistic PDES.

\item We implement asynchronous one-sided communication, increasing computation-communication overlap and eliminating handshake latency.

\item We develop adaptive message aggregation and routing strategies within a 2D-HyperX virtual topology to optimize bandwidth utilization for small message traffic.

\item We utilize up to 174 compute nodes of an HPC cluster to simulate a multi-area SNN model of the marmoset brain, achieving up to 1.4$\times$ speedup.

\end{itemize}

The remainder of this paper is structured as follows. \Cref{sec:background} provides the background and related work on SNN synchronization and communication. \Cref{sec:rollback} details the rollback mechanism. \Cref{sec:async} outlines the asynchronous communication utilizing UNR. \Cref{sec:routing} introduces our strategy of message routing, aggregation, and process mapping. \Cref{sec:experiment} presents the experimental studies. \Cref{sec:conclusion} and \Cref{sec:future} summarize our contributions and prospects for future work.

\begin{figure}[t]
  \centering
  \begin{subfigure}[b]{0.31\textwidth}
    \includegraphics[width=\textwidth]{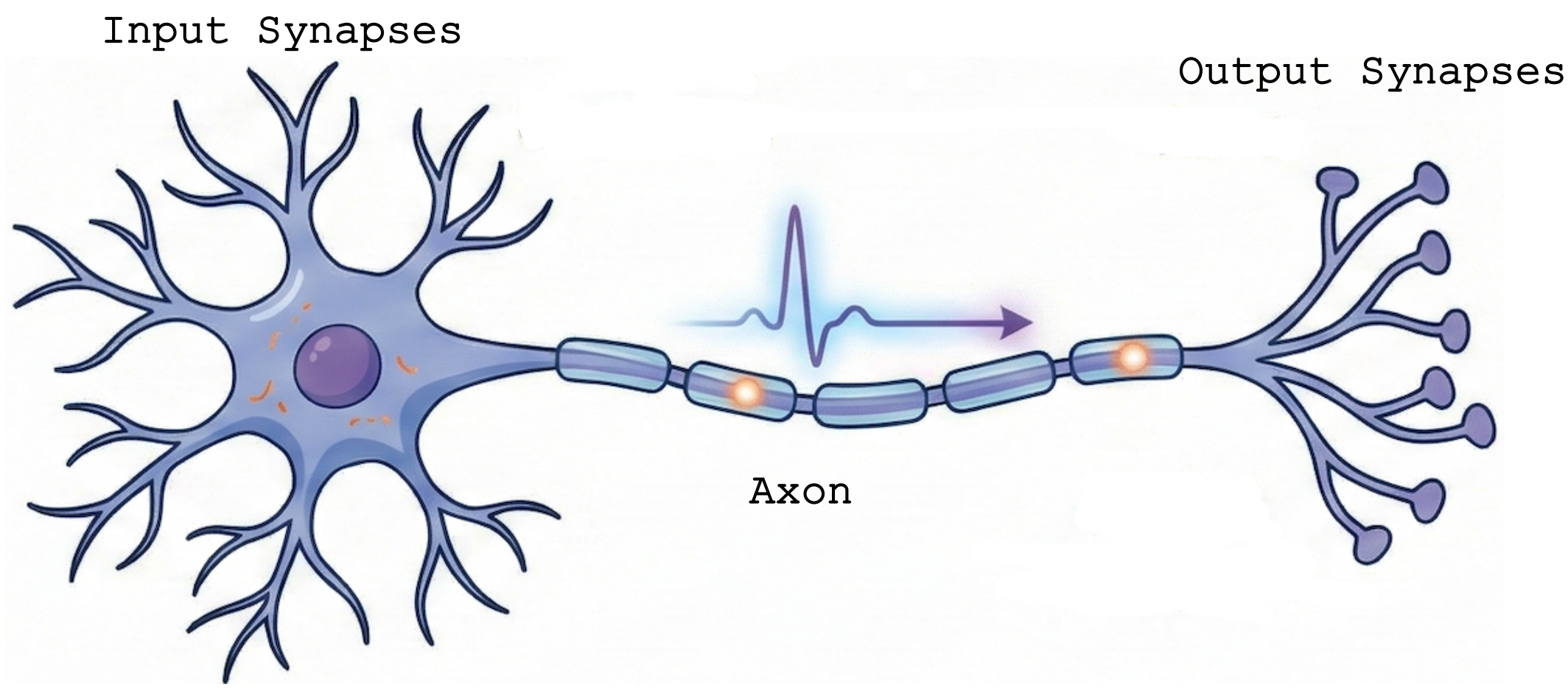}
    \caption{Neuron}
  \end{subfigure}
  \begin{subfigure}[b]{0.17\textwidth}
    \includegraphics[width=\textwidth]{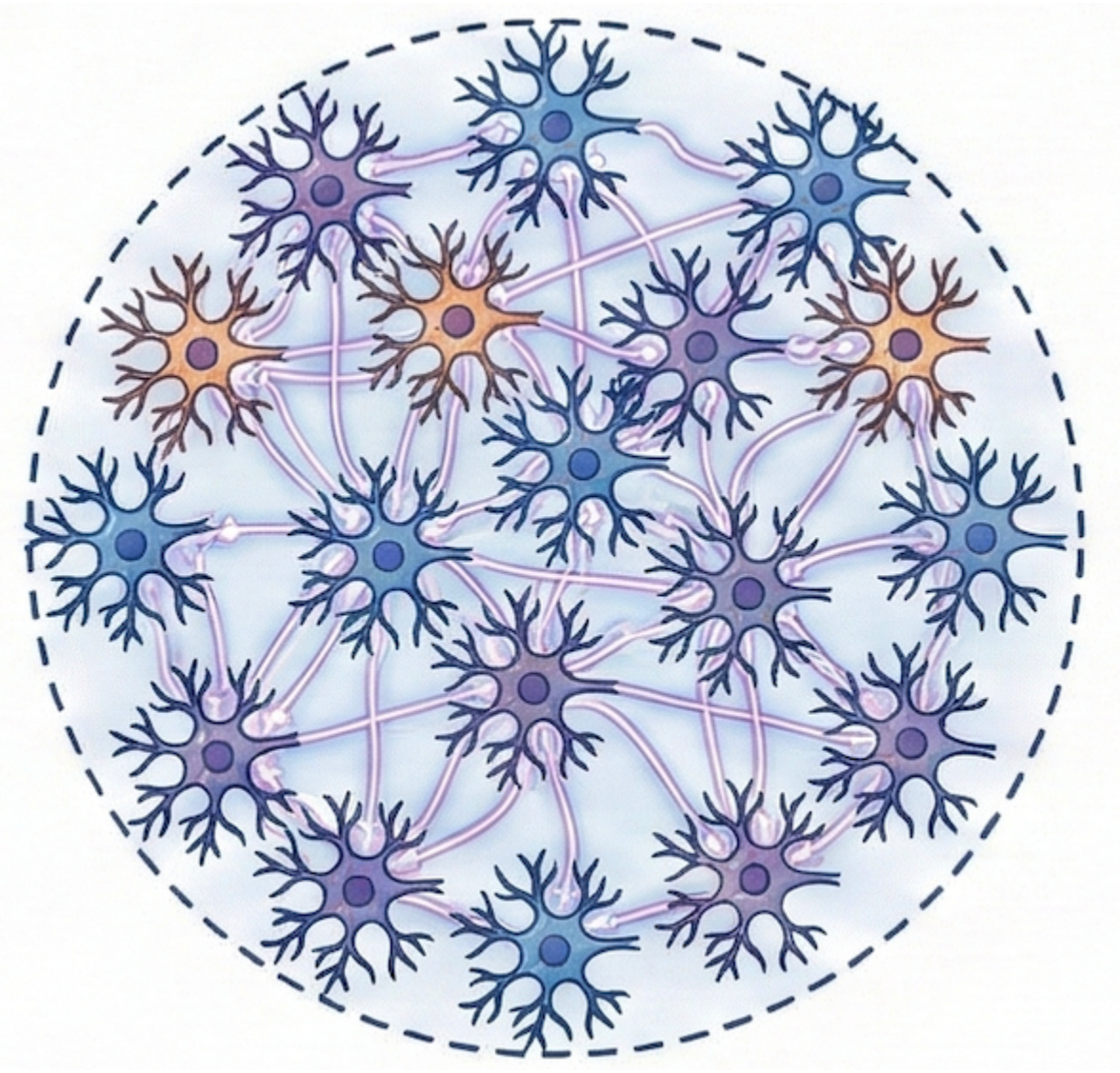}
    \caption{Population}
  \end{subfigure}
  \begin{subfigure}[b]{0.28\textwidth}
    \includegraphics[width=\textwidth]{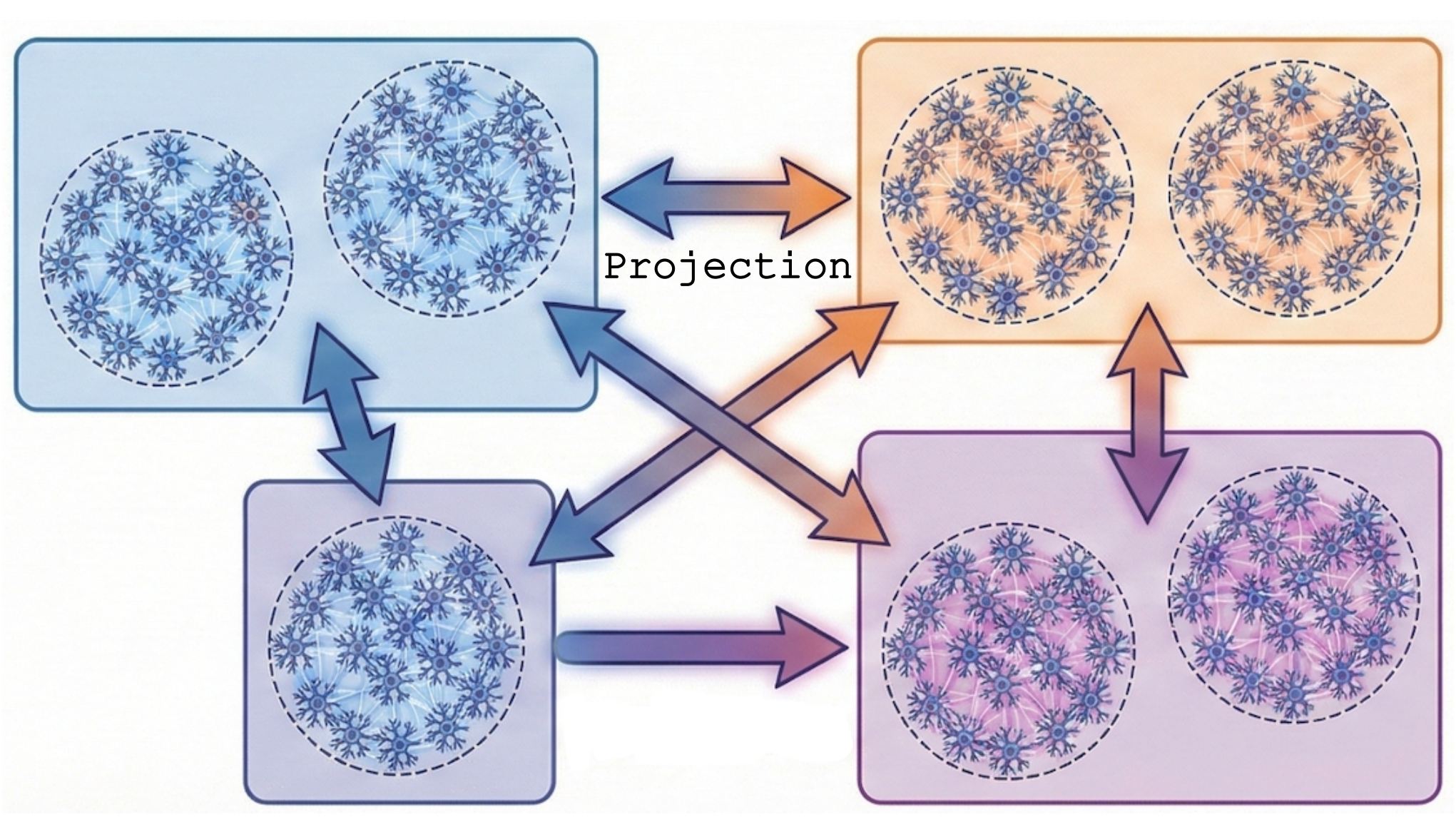}
    \caption{Area}
  \end{subfigure}
  \begin{subfigure}[b]{0.20\textwidth}
    \includegraphics[width=\textwidth]{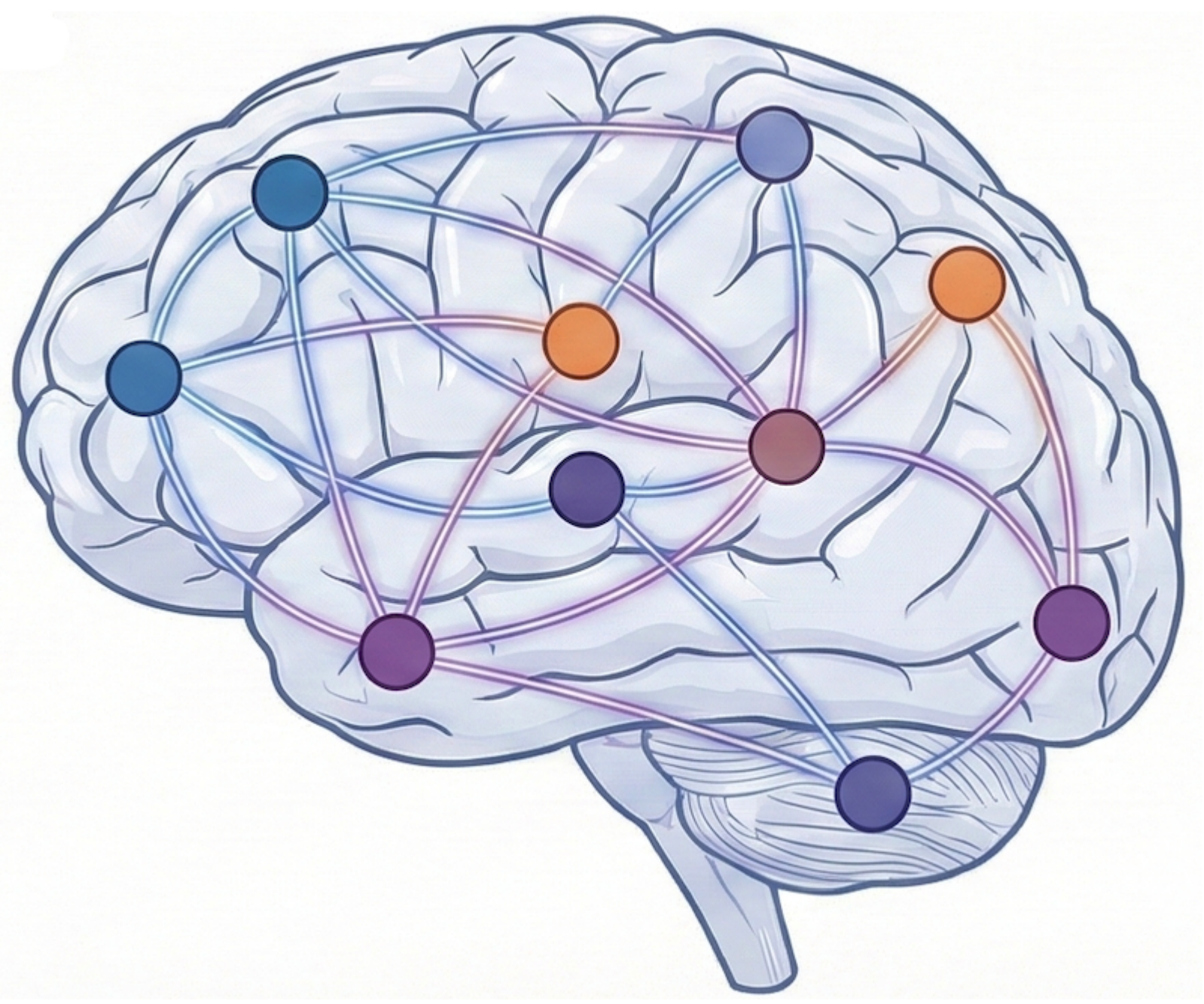}
    \caption{Brain}
  \end{subfigure}
    \caption{The components of the SNN model. (a) Neurons handle and generate spikes, which are transmitted by synapses. (b) Neurons with similar biological features are grouped together to form populations. (c) Areas are composed of many populations and are responsible for brain functions. (d) The brain comprises multiple interconnected areas, forming a complex network structure.}
    \label{fig:struct}
\end{figure}

\section{Background and Related Work}
\label{sec:background}

This section provides the necessary background to contextualize our work.
\Cref{sec:SNNsim} introduces the foundational SNN model, explaining why synchronization and communication become bottlenecks in large-scale simulations.
We then review existing SNN simulators, categorized into synchronization strategies (\Cref{sec:background_sync}) and communication mechanisms (\Cref{sec:background_comm}).

\subsection{SNN Simulation}
\label{sec:SNNsim}

The SNN model can be depicted in a hierarchical view, as illustrated in \Cref{fig:struct}. Neurons and synapses are two basic components of the SNN. Neurons, the fundamental building blocks of SNNs, mimic the functionality of biological neurons. They process incoming spike signals and determine whether to generate output spikes based on changes in their membrane potential. Synapses serve as the communication channels between neurons, facilitating the transmission of spike signals. Each synapse has a weight that modulates the signal's strength.

From a more macroscopic perspective, the SNN model is composed of populations and projections \cite{davison2009pynn}. Populations are made up of groups of neurons that share similar properties. Projections define the synaptic connections between these populations. Brain areas, which are larger functional units composed of multiple distinct populations, correspond to the diverse functional regions in biological brains, such as the visual cortex. These areas are also interconnected through synapses, forming a complex network structure that enables the brain to carry out advanced information-processing tasks. SNN simulations involve detailed modeling of the interactions between neurons, populations, and brain areas via synapses, as well as the monitoring of inputs and outputs.

In large-scale SNN simulations, the key challenges arise primarily from frequent synchronization and irregular communication patterns, which can profoundly impact the parallel efficiency of simulations \cite{zenke2014limits, fernandez2019communication}.
The need for frequent synchronization stems from the fact that, within each effective refractory period, every neuron must confirm the arrival of all spike signals sent by its input synapses in the previous period before it can proceed, effectively serializing the computation across dependent processes.
Meanwhile, the communication pattern is inherently irregular because only a small fraction of neurons fire spikes in a period, and the specific target neurons connected via output synapses differ for each sender, leading to unpredictable and fine-grained communication.
These application characteristics collectively result in severe performance bottlenecks, manifesting as frequent processor idle time due to synchronization waits and a flood of small, inefficient network packets that underutilize bandwidth.
Therefore, existing works have conducted extensive research on synchronization and communication issues in SNN simulation.

\begin{figure*}[ht]
  \centering
  \begin{subfigure}[b]{\linewidth}
    \includegraphics[width=\linewidth]{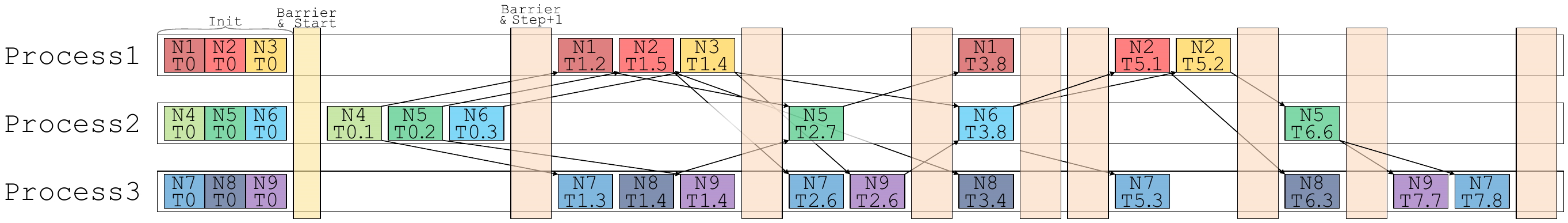}
    \caption{Time-Driven SNN Simulation: Synchronize after each time step.}
    \label{fig:Time_vs_Event_1}
  \end{subfigure}
  \begin{subfigure}[b]{\linewidth}
    \includegraphics[width=\linewidth]{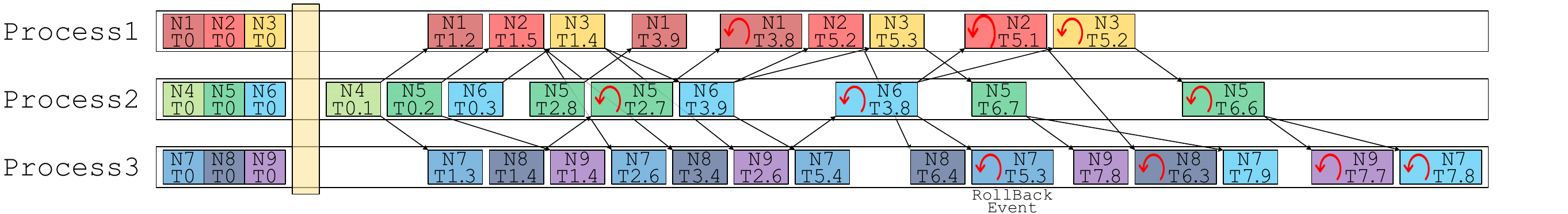}
    \caption{Optimistic Parallel Distributed Event SNN Simulation: No synchronization and rollback when errors occur.}
    \label{fig:Time_vs_Event_2}
  \end{subfigure}
  \begin{subfigure}[b]{\linewidth}
    \includegraphics[width=\linewidth]{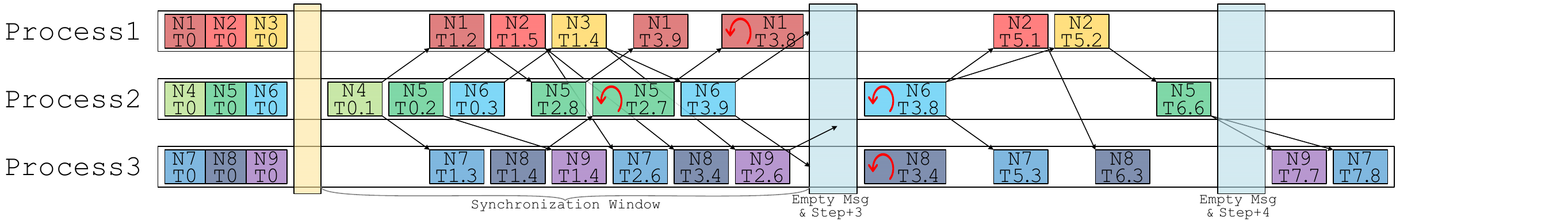}
    \caption{Optimistic Hybrid SNN Simulation (ours): Synchronize every $k$ steps and rollback within each window.}
    \label{fig:Time_vs_Event_3}
  \end{subfigure}
  \caption{
    Synchronization Mechanisms in Time-Driven, Event-Driven, and OACM SNN Simulation.
    \texttt{N\#} represents \texttt{Neuron\#}, and \texttt{T\#} represents the spike that arrives at \texttt{Virtual Time\#}.}
  \label{fig:Time_vs_Event}
\end{figure*}

\subsection{Synchronization Mechanisms in Time-Driven and Event-Driven SNN Simulation}
\label{sec:background_sync}

There are two primary methods for large-scale SNN simulation: time-driven and event-driven simulation.
The most evident difference between the two simulation methods is how the simulation time advances, which leads to different synchronization mechanisms. 

Time-driven simulators operate by advancing through fixed timesteps, which must be set smaller than the effective refractory period to ensure accurate neural dynamics, thus requiring frequent synchronization and communication as illustrated in \Cref{fig:Time_vs_Event_1}.
Each simulation step is typically divided into two distinct phases.
In the neuron computation phase, all active neurons (i.e., not in a refractory state) integrate incoming signals and update their internal states based on the neuron model, deciding whether to emit a spike to post-synaptic neurons.
In the spike propagation phase, spikes generated in the current or preceding steps are transmitted to their target synapses, taking synaptic delays into account.
Synapses then update their states and deliver the resulting inputs to the post-synaptic neurons.
Generally, there is an explicit or implicit global synchronization in the spike propagation phase.
Currently, time-driven simulators dominate the field due to their ease of implementation \cite{Gewaltig:NEST, plesser2007efficient, goodman2008brian, migliore2006parallel, zenke2014limits, yavuz2016genn, ahmad2018spike, chou2018carlsim, stimberg2019brian, igarashi2019large, fernandez2019communication, golosio2020toward, qu2020high, golosio2021fast, bautembach2021multi, niedermeier2022carlsim, liu2022regularizing, du2022low, tiddia2022fast, lu2023digital, lyu2024cortex}, despite incurring substantial synchronization overhead as the simulation scale increases.

On the other hand, event-driven simulators offer arbitrary time resolution and advance the simulation via variable time increments.
They only trigger updates upon the arrival of specific events, such as spike exchanges, eliminating the need for fixed time steps and reducing computational overhead by only updating relevant components.
However, conservative event-driven SNN simulation requires an event to be processed if and only if all of its earlier events have been processed, which leads to synchronization between connected neurons to ensure correctness.
Thus, most event-driven SNN simulators are limited by their reliance on specialized neuromorphic hardware \cite{zhang2019asynchronous, lee2021neuroengine, lee2022neurosync}, a lack of multi-threading simulation support \cite{peng2018novel, ros2006event, mo2022evtsnn, pecevski2014nevesim}, or no support for Parallel Distributed Event Simulation (PDES) \cite{mo2021edha, bautembach2021even, bautembach2020faster, susi2018fns}, which restricts their versatility and scalability.
To improve scalability, the simulators of \cite{plagge2018nemo, pimpini2022speculative} utilize optimistic PDES to relax the synchronization condition under which incorrect event computation can be rolled back.
Nevertheless, the heterogeneous activity patterns across different brain regions lead to inter-process load imbalance, which may result in a large number of secondary rollbacks, as shown in \Cref{fig:Time_vs_Event_2}.

The simulator of \cite{naverosSpikingNeuralSimulator2015} combines event-driven and time-driven strategies.
The low-activity parts of the neural network can be simulated on CPU using event-driven methods, while the high-activity subsystems can be simulated on either CPU or GPU using time-driven methods.
In contrast, OACM seamlessly integrates both strategies by leveraging sparse spiking activity and bounded synaptic delay.

\subsection{Communication Implementation in SNN Simulation}
\label{sec:background_comm}

At present, most synchronization mechanisms in SNN simulators are implemented by MPI communication operations.
Some simulators employ synchronous MPI functions, such as point-to-point communication (e.g., MPI\_Send and MPI\_Recv) \cite{tiddia2022fast, qu2023enlarge, du2024hrcm} and collective communication (e.g., MPI\_Alltoall and MPI\_Allgatherv) \cite{Gewaltig:NEST, zenke2014limits, qu2023enlarge, lyu2024cortex}.
These functions inherently guarantee the delivery of messages and the synchronization of all processes, thereby obviating the need for additional explicit synchronization measures.
The other simulators \cite{igarashi2019large, fernandez2019communication, du2022low, tiddia2022fast} incorporate asynchronous communication protocols to achieve synchronization.
These simulators harness asynchronous communication operations, such as MPI\_Isend/MPI\_Irecv, in conjunction with MPI\_Request to track communication requests.
Subsequently, synchronization is implemented by MPI\_Waitall to confirm message arrival and to ensure data consistency across processes.
While the former streamlines implementation by integrating synchronization within communication functions, the latter gains enhanced flexibility and control, especially in situations that demand meticulous coordination or the overlapping of computation with communication.


As for communication optimization, overlapping communication with computation is a common approach: 
CORTEX \cite{lyu2024cortex} optimizes communication by using a dedicated thread for communication in each process to overlap communication and computation. This approach allows synaptic interactions to be handled independently of the computation threads, enabling efficient dataflow and spike propagation without dependencies at both the process and thread levels.
MONET \cite{igarashi2019large} utilizes a tile partitioning method and asynchronous point-to-point communication, where the communication interval is set to half of the minimum transmission delay, thus enabling the computation to overlap the asynchronous communication of MPI\_Isend/MPI\_Irecv.
DTB \cite{lu2023digital} optimizes communication by employing multiple threads for parallel computation and communication, with GPUs managing state variables and spike sequences to enhance efficiency. It also implements a method to prevent communication contention by prioritizing the order of data transmission and using a sending thread to manage information flow, ensuring smooth and efficient data exchange. Other optimization techniques, such as message aggregation and routing, have also been applied in some studies.
HRCM \cite{liu2022regularizing, du2024hrcm} proposes a virtual communication topology (VCT), which organizes processes into an n-dimensional virtual topology and uses multi-stage message merge-forward to reduce the total number of messages transmitted.
Du et al. \cite{du2022low} place neurons with strong communication demands on the same GPU and use two-level routing to optimize communication, where they cluster GPUs into groups and minimize inter-group connections.
Carlos et al. \cite{fernandez2019communication} employ hypergraph-based partitioning to minimize inter-process communication by carefully placing neurons and introducing neighborhood exchange algorithms to overlap the communication handshake with the send-receive data phase.

Communication, along with synchronization, has long been recognized as the most time-consuming part of distributed SNN simulations \cite{zenke2014limits, fernandez2019communication, cremonesi2020understanding}.
To tackle the synchronization and communication bottleneck, we propose a more versatile approach that does not depend on the minimum synaptic delay and incorporates communication optimizations that further minimize the synchronization frequency and enhance the communication efficiency of simulations.
The subsequent sections will provide a detailed exposition of our strategies.

\section{Optimistic Hybrid SNN Simulation}
\label{sec:rollback}

\subsection{Design Overview}

Optimistic Hybrid SNN Simulation seamlessly integrates time-driven simulation and Optimistic PDES. It inherits the implementation simplicity of time-driven simulators while eliminating the need for frequent global synchronization. Simultaneously, it strategically applies Optimistic PDES techniques locally, thereby avoiding the extensive secondary rollbacks caused by load imbalance.

The core design of Optimistic Hybrid SNN Simulation involves performing global synchronization every $n$ timesteps, which we refer to as a ``synchronization window'', as shown in \Cref{fig:Time_vs_Event_3}. By adopting this approach, the total number of synchronizations is reduced by a factor of $n$. To ensure simulation correctness, we incorporate the rollback strategy from Optimistic PDES, but confine rollbacks within each synchronization window. The pseudocode for the entire rollback procedure is presented in \Cref{algo:rollback}.

\begin{table}[b]
  \caption{Taxonomy of Spike Types}
  \label{tab:spike}
  \centering
  \begin{tabular}{|c|c|c|c|}
    \hline
    Types & Condition & Action & Rollback\\
    \hline
    Timely spike & $e \geq t$ & Add & No \\
    Timely anti-spike & $e \geq t$ & Sub & No\\
    Late spike & $e < t$ & Add & Yes\\
    Late anti-spike & $e < t$ & Sub & Yes\\
    \hline
\end{tabular}
\end{table}

\begin{algorithm}[t]
\caption{Rollback Procedure}
\label{algo:rollback}
{\bf Input:} $t$: current timestep, $B[]$: received spike buffer, \\
$R[]$: rollback step of neurons, $D[]$: synaptic delay
\begin{algorithmic}[1]
\State // Handle spikes
\For{each spike $i$ in buffer $B$} 
    \For{each target neuron $j$}
        \State $e$ = $B[i].genTime$ + $D[B[i].src][j]$
        \State Write to neuron $j$'s input at timestep $e$
        \If{$e < t$}
            \State $R[j]$ = $t$ - $e$
        \EndIf
    \EndFor
\EndFor
\State // Update neurons
\For{each neuron $i$} 
    \State Set the neuron to the state of timestep $(t-R[i]-1)$
    \For{$j$ from $R[i]$ to $1$} 
        \State Record whether neuron spikes before update.
        \State Update and save neuron state of timestep $(t-j)$
        \State Record whether neuron spikes after update.
        \If{Spiked before update but not after}
            \State Generate anti-spike with timestep $(t-j)$
        \ElsIf{Spiked after update but not before}
            \State Generate spike with timestep $(t-j)$
        \EndIf
    \EndFor
    \State Update and save neuron state of timestep $t$
    \State Record whether neuron spikes after update.
    \If{Spiked after update}
        \State Generate spike with timestep $t$
    \EndIf
\EndFor
\end{algorithmic}
\end{algorithm}

\subsection{Implementation}

To begin with, we establish a clear taxonomy of spike types, listed in \Cref{tab:spike}. Central to this categorization is the concept of the ``effective timestep'' $e$. This term refers to the exact timestep at which the synaptic input, triggered by a spike, affects the post-synaptic neurons. The effective timestep is determined by adding the spike's generation timestep $g$ to the synaptic delay $d$. With this concept in mind, spikes can be classified by comparing $e$ with the current timestep $t$, and by determining if they are anti-spikes:

\begin{itemize}
\item \textbf{Timely spike}: Occurs when $e \geq t$. Processed immediately if timestamps match, otherwise queued for future processing.

\item \textbf{Late spike}: Arises when $e<t$, typically due to network delays or rollback recalculations from other processes. Triggers rollback operations.

\item \textbf{Timely anti-spike}: Used to neutralize erroneous spikes that are generated before rollback. Occurs when $e \geq t$. Processed immediately or queued.

\item \textbf{Late anti-spike}: Functions similarly to Timely anti-spike. Occurs when $e<t$ and necessitates rollback operations.

\item \textbf{Straggler spike}: Refers to both Late spike and Late anti-spike, which can cause rollbacks.

\end{itemize}

Next, we manage the rollback state of each neuron in a fine-grained manner, rather than employing uniform control over the entire process. Each neuron is associated with a circular array whose length matches the size of the synchronization window. This array records the state of the neuron at every timestep within the specified synchronization window, encompassing parameters such as membrane potential, refractory period count, and synaptic input. Such fine-grained storage enables us to perform rollbacks for individual neurons rather than for all neurons within a process, avoiding unnecessary repeated computation. 
Since only a subset of neurons undergoes rollback, the recalculation for timestep $i+1$ is significantly shortened. Here, for implementation, we employ the checkpoint technique rather than incremental computation. This decision is predicated on the fact that the dynamic equations used for calculating neuron states do not support incremental methods.

Then, we design a mechanism to handle different spike message types. Upon receiving spikes, a process first calculates the effective timestep $e$ for each target neuron using the spike's generation time $g$ and synaptic delay $d$, and then modifies that neuron's input buffer at timestep $e$. For an anti-spike, the corresponding value is subtracted from that neuron's input buffer; otherwise, the value is added. If $e<t$, that neuron's rollback step count $r$ is set to $t-e$. During neuron state updates, a non-zero $r$ triggers the rollback procedure: the neuron's state is reset to the state at $t-r-1$, and then the state is recalculated and saved step by step from $t-r$ onward. If the spike emission during the rollback does not match the history, new spikes or anti-spikes will be generated. Finally, the state at $t$ is calculated, and it is checked whether a new spike is generated. All generated spikes are buffered and transmitted later.

Finally, the synchronization mechanism is further complicated by the introduction of rollbacks. Upon reaching the synchronization timestep $s$ at the end of each synchronization window, a process halts and verifies the receipt of all inter-process messages before updating neuron state at $s$. Spikes generated during updates at $s$ are retained for transmission after synchronization. The neuron update procedure calculates a local maximum rollback count $r_{max}$ in that process, and then computes the global maximum rollback count $R_{max}$ via an MPI\_Allreduce operation. Next, each process reverts all the states to timestep $s - R_{max} - 1$ and resumes simulation from timestep $s - R_{max}$. Synchronization serves to purge historical states stored prior to timestep $s - R_{max}$ and to set an upper limit on rollback, preventing states from being reverted further than timestep $s - R_{max}$. 

\subsection{Design Analysis}

This implementation offers three key advantages:
First, ease of implementation is crucial for SNN simulation, as SNN models frequently require adjustments. Using PDES for large-scale SNN simulation would require domain experts to understand complex concepts such as LP, KP, GVT, and fossil collection, whereas time-driven implementations are more comprehensible. This explains why the vast majority of known simulators employ time-driven approaches.
Second, the synchronization window imposes an upper bound on the historical states that need to be stored, thereby avoiding the massive memory consumption issues in traditional Optimistic PDES under load imbalance. This is particularly important for SNN simulations, which typically require substantial memory to store connection structures.
Finally, this approach eliminates the overhead of precise event time ordering, requiring only bucket sorting as in time-driven simulators, rather than the exact event ordering required in Optimistic PDES.

\section{Asynchronous One-Sided Communication Model}
\label{sec:async}

\subsection{Communication Requirements and Existing Limitations}
\label{sec:async_requirements}

Communication latency and computation-communication overlap are critical factors for SNN simulation performance. 
As introduced in \Cref{sec:rollback}, our Optimistic Hybrid SNN Simulation still relies on rollback operations within each synchronization window to ensure correctness, where communication delays directly impact whether spike data can arrive in a timely manner between processes. 
Achieving complete computation-communication overlap is therefore essential to maximize simulation efficiency.

Existing approaches typically employ MPI-based asynchronous communication, such as two-sided non-blocking operations (MPI\_Isend/MPI\_Irecv) or one-sided operations (MPI\_Put/MPI\_Get). 
However, these implementations suffer from implicit communication overhead.
Two-sided non-blocking operations require a handshake between sender and receiver (Rendezvous Protocol), or an additional memory copy (Eager Protocol).
One-sided operations require pre-synchronization to ensure buffer or data readiness, and post-synchronization to notify the peer that data is ready. 
These synchronization operations are typically performed through MPI\_Win\_Fence, consuming at least a single trip time.
Note that the synchronization referred to in this section represents the synchronization between sender and receiver, rather than the global barrier synchronization operation referred to in previous \Cref{sec:rollback}.

\subsection{Implementation}
\label{sec:async_implementation}

To overcome these limitations, we leverage the Unified Notification RDMA Library (UNR) \cite{feng2024unr}.
UNR provides one-sided RDMA Read and RDMA Write operations like MPI, 
but it utilizes low-level networking programming interfaces on different platforms to obtain the transmission completion and data arrival signals of these operations, thus avoiding manual post-synchronization.
These signals could be automatically placed into a queue by UNR, enabling efficient data processing.
As for pre-synchronization, UNR also provides a paradigm that hides it within preceding communications.

Building upon UNR's capabilities, we implement an optimized asynchronous one-sided communication pattern specifically designed for SNN spike data transfer within each synchronization window.
The spike data are directly written into the receiver buffer using the one-sided RDMA Write operation provided by UNR.
The pre-synchronization and post-synchronization can be hidden or avoided in our communication model.

As for pre-synchronization, we use a dedicated buffer for each sender process within the current synchronization window; thus, all the pre-synchronization is hidden in the global barrier operations among synchronization windows.
By isolating buffer access per sender process, multiple source processes can write to the same destination process concurrently without requiring coordination or synchronization between senders.
In addition, there is a buffer size upper bound for each sender process since there is a global barrier operation after each synchronization window.
The buffer only needs to cache spike data for each step within a synchronization window.
A typical total receive buffer size for each process is 256MB in our experiments, which is relatively acceptable.
Although this strategy seems to lack scalability, the biological characteristics of SNNs dictate that when the number of neurons simulated by a process is fixed, the number of pulse signals received will not increase as the total simulation scale grows, and the virtual topology introduced in \Cref{sec:routing} further reduces the buffer used for redundancy.

\begin{figure}[t]
  \centering
  \begin{subfigure}[b]{\linewidth}
    \includegraphics[width=\linewidth]{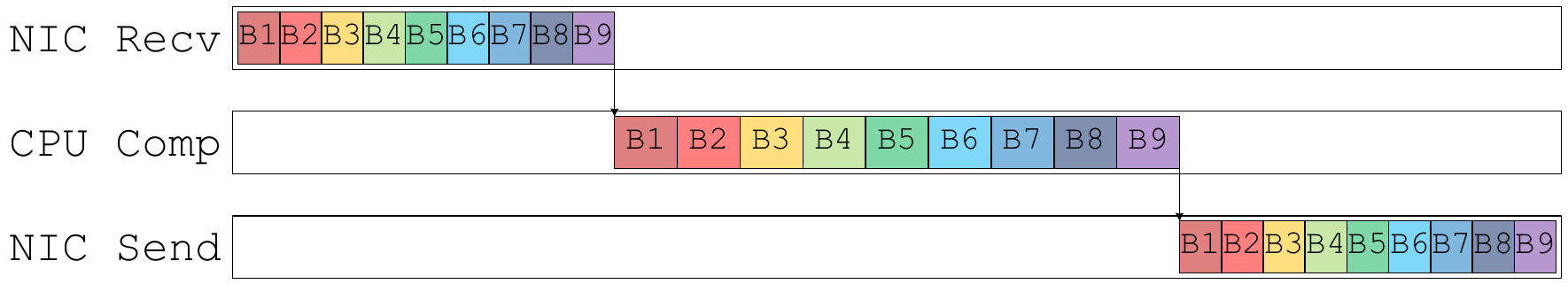}
    \caption{Serial Computation-communication.}
  \end{subfigure}
  \begin{subfigure}[b]{\linewidth}
    \vspace{10pt}
    \includegraphics[width=\linewidth]{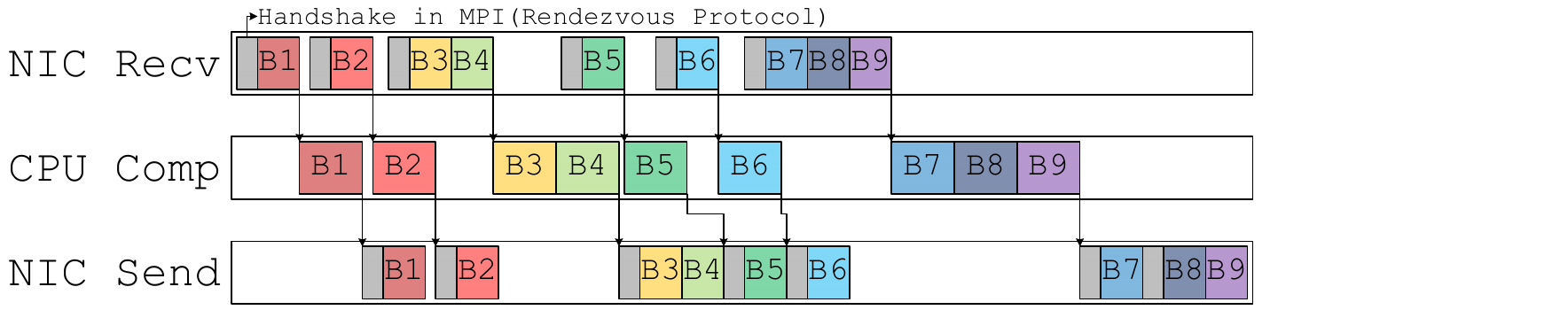}
    \caption{Computation-communication overlap with MPI.}
  \end{subfigure}
  \begin{subfigure}[b]{\linewidth}
    \vspace{10pt}
    \includegraphics[width=\linewidth]{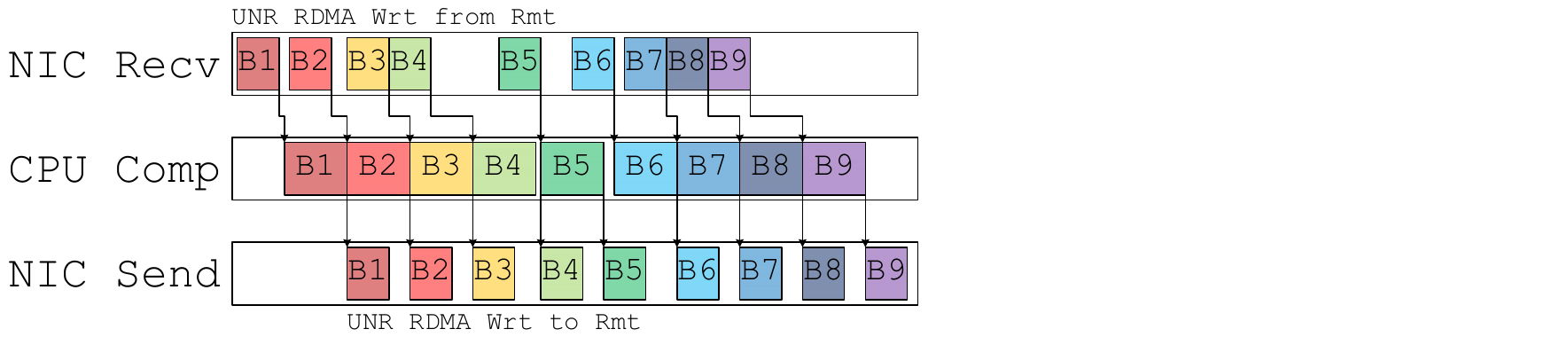}
    \caption{Computation-communication overlap with UNR (ours).}
  \end{subfigure}
    \caption{Computation-communication overlap mechanisms}
    \label{fig:overlap}
\end{figure}

As for post-synchronization, we directly leverage UNR to obtain the arrival signals of spike data.
Based on timely message arrival notifications provided by UNR, we implement a progressive communication truncation mechanism as shown in \Cref{fig:overlap}.

Upon detecting the arrival of  a message containing some spike data, the process proactively terminates the communication phase without waiting for the arrival of future messages, enabling the complete overlap of computation and communication.
The effectiveness and correctness of this approach are guaranteed by our rollback mechanism introduced in the previous section.
When messages are not received in time, the simulator will trigger the rollback mechanism to reprocess the corresponding spikes and state updates. So there is no need to worry about the causal correctness of the simulation results.

\begin{figure}[t]
  \centering
      \includegraphics[width=\linewidth]{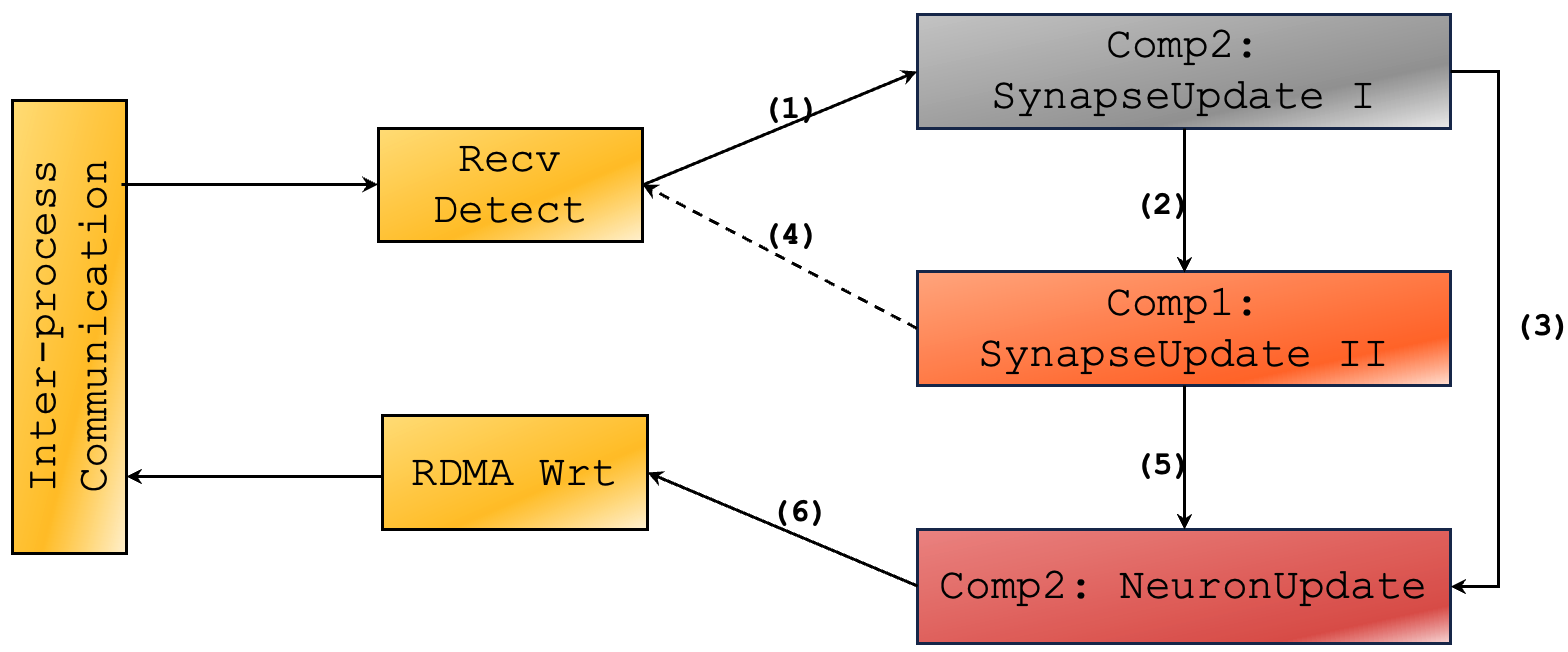}
    \caption{The procedure and dataflow of the simulation. (1) The newly received spikes initially undergo `SynapseUpdate I'. (2) Those spikes with an effective timestep exceeding the next timestep are reserved for `SynapseUpdate II' to overlap communication. (3) Those in-time spikes are immediately processed to accumulate synapse weights into neuron inputs. (4) Upon completion of `SynapseUpdate II', the system stops detecting incoming messages. (5) Those reserved spikes are computed and added to neuron inputs once they reach the corresponding timestep. (6) Spikes generated during `NeuronUpdate' are transmitted via UNR RDMA write operations, concluding a timestep loop. }
    \label{fig:procedure}
\end{figure}

The overall simulation procedure and dataflow are illustrated in \Cref{fig:procedure}. The key mechanism is the `SynapseUpdate II' phase, which reserves spikes whose effective timestep has not yet arrived so that they can be processed during communication, thereby enabling full computation-communication overlap.

\subsection{Design Analysis}

Our Asynchronous One-Sided Communication Model essentially uses RDMA-based one-sided communication and eliminates pre-synchronization and post-synchronization at the cost of additional memory consumption. Since UNR provides fine-grained data arrival information, it enables computation-communication overlap at fine granularity.

\begin{figure*}[ht]
  \centering
  \begin{subfigure}{\linewidth}
    \centering
    \includegraphics[width=0.8\linewidth]{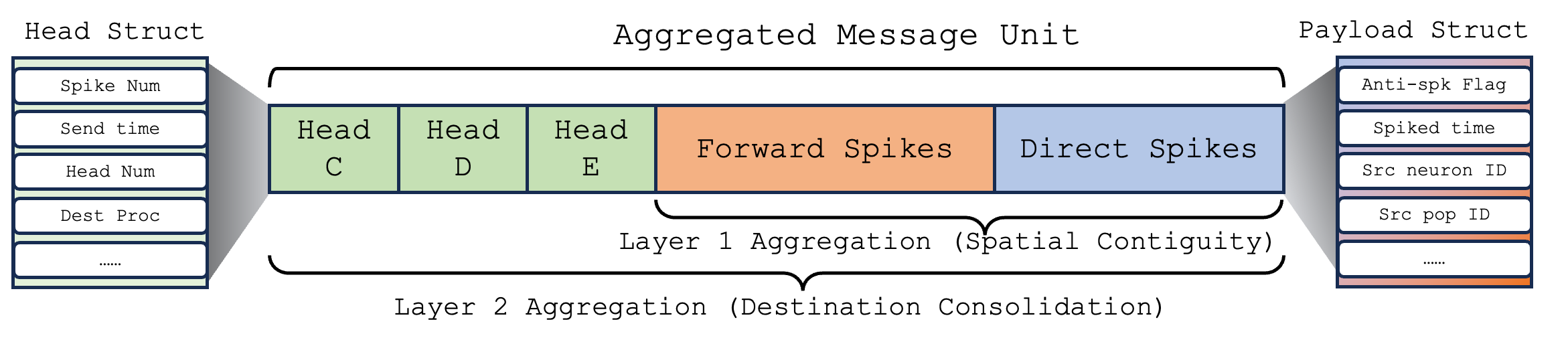}
    \subcaption{Data Structures and Aggregation Layers.}
  \end{subfigure}\\[6pt]
  \begin{subfigure}{0.2\linewidth}
    \centering
    \includegraphics[width=\linewidth]{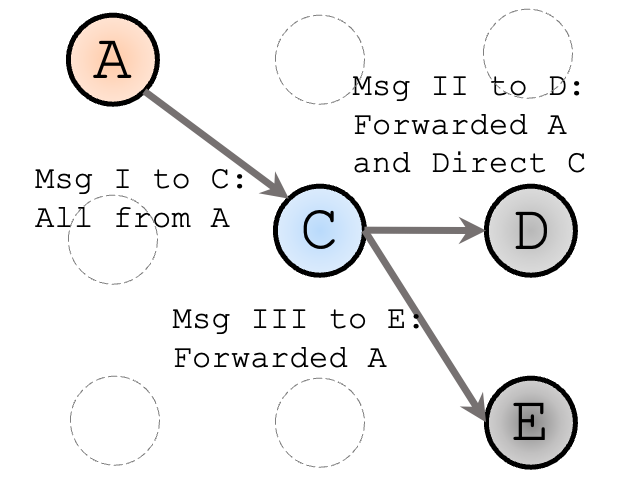}
    \subcaption{Routing Example.}
  \end{subfigure}\hfill
  \begin{subfigure}{0.78\linewidth}
    \centering
    \includegraphics[width=\linewidth]{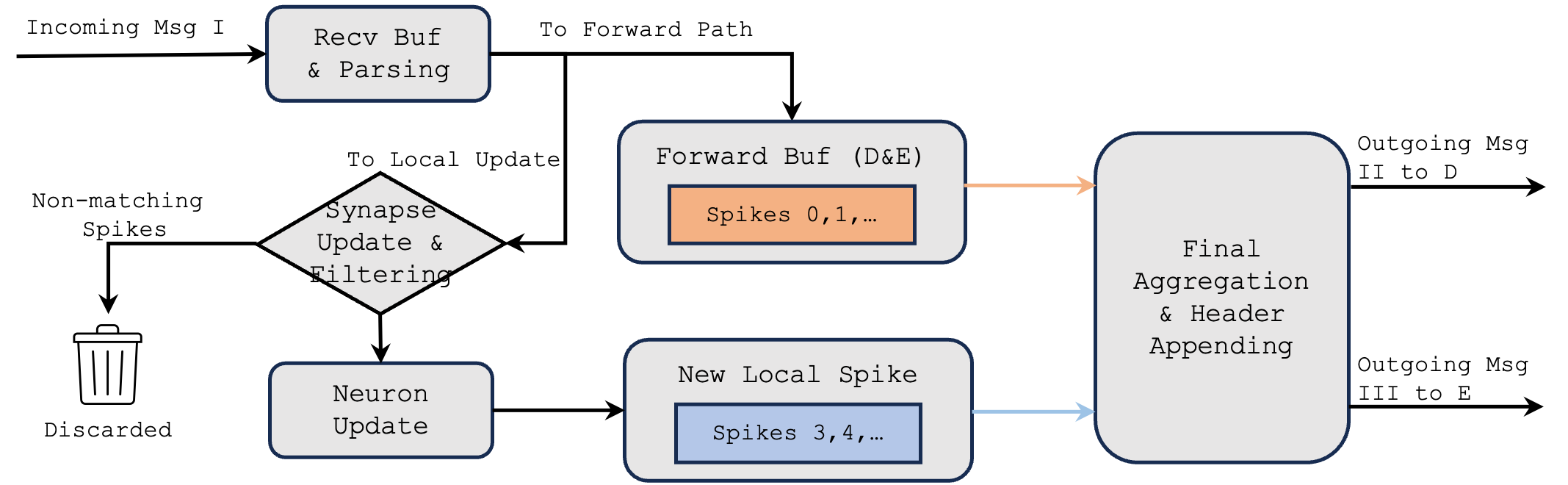}
    \subcaption{Process C Internal Lifecycle.}
  \end{subfigure}
  \caption{Message routing and aggregation. Process A sends message I to C. Its content is also intended for D and E. Upon receiving, C stores the relevant spikes and discards the irrelevant ones. Then, C sends message II to D and message III to E, both of which include the spikes forwarded from A and the spikes generated within C.}
  \label{fig:message}
\end{figure*}

\section{Message Routing and Aggregation}
\label{sec:routing}

\subsection{Routing on Virtual Topology}

Although UNR offers high-performance point-to-point communication, it does not provide any collective communication primitives.
The existing MPI collective communication is inefficient and introduces extra global synchronization overhead, making it impossible to employ our aforementioned optimizations and failing to satisfy the requirements for irregular communication in SNN simulations.
By contrast, naive point-to-point communication among all processes involves a large number of small data packets and its quadratic scaling complexity in the communication phase.

To optimize bandwidth utilization and enhance communication efficiency, we leverage dimension-order routing in a virtual 2D-HyperX topology consisting of all processes.
This approach significantly reduces the number of communication connections and small data packets while preserving functional equivalence to all-to-all message passing.

We logically organize $N$ processes into a 2D-HyperX of size $(\lfloor \sqrt{N} \rfloor, \lceil \sqrt{N} \rceil)$, balancing row and column dimensions.
Each process is uniquely identified by its coordinate.
Communication between processes is restricted to those sharing the same X or Y dimension.
When process $a$ sends a message to process $b$, it first checks for direct adjacency. If $b$ is not adjacent, the message is sent to process $c$, which shares the same $y$ coordinate with $b$ and the same $x$ coordinate with $a$. Process $c$ then forwards the message horizontally to $b$ along the X-axis. This is known as YX routing, and the reason for not using XY routing will be explained in \Cref{sec:mapping}.
This orthogonal decomposition ensures that any inter-process communication requires at most two hops, reducing the number of communications per process from $O(N)$ to $O(\sqrt{N})$ compared to the naive all-to-all pattern.

While routing in a virtual topology is common in HPC applications, it is seldom used in SNN simulations and our work surpasses prior efforts. 
Earlier GPU cluster-based large-scale SNN simulations \cite{liu2022regularizing, du2024hrcm} have adopted the 2D-HyperX virtual topology. Their results indicate that 2D routing optimally balances communication traffic and runtime performance.
Going further, our approach differs from VCT in three key ways:
(a) We use the UNR library for asynchronous one-sided communication, avoiding the synchronous MPI\_Send/MPI\_Recv calls.
(b) They split communication into 2 stages and wait for all messages before proceeding, introducing extra synchronization overhead. In contrast, our method sends messages immediately upon spike arrival or generation, incorporating a rollback mechanism to reduce synchronization overhead.
(c) Our approach considers communication heterogeneity and includes process mapping as detailed in \Cref{sec:mapping}.

\subsection{Message Aggregation}


To further optimize inter-process communication efficiency and enhance bandwidth utilization, we implement a two-layer message aggregation mechanism.
The structure of an example message is illustrated in \Cref{fig:message}.
Each message includes two types of components: a header and a spike payload.
The header contains multiple metadata entries, each corresponding to a destination process and encoding essential parameters, including destination ID, the address of the next header, and the number of spikes.
The spike payload is categorized into two types of spikes: forward spikes, which originate from other processes and are routed by the current process, and direct spikes, which are locally generated to influence processes on the same axis or to be forwarded through them.

Rather than employing the conventional approach of separately filling in specific spikes for each destination process, all direct spikes pertaining to the same brain area are aggregated into a unified spike payload.
Subsequently, destination-specific headers are appended, thereby maximizing message coalescence. Upon receiving a message, a process first parses header metadata and distributes spike payload to the corresponding buffers based on the destination identifiers.
During the synapse update, if a spike does not match any post-synaptic neuron resident in the process, then it will be discarded. This filtering occurs prior to synaptic weight evaluation, thereby eliminating unnecessary computation cycles for non-local interactions.

Thus, a two-layer aggregation is established:
The first layer combines direct and forwarded spikes into a single transmission unit, exploiting spatial contiguity of target processes along topology dimensions.
The second layer consolidates spikes generated by the same source neuron to multiple destinations, distributing the filtering burden to destination processes. 

\subsection{Process Mapping}

\label{sec:mapping}

When simulating large-scale brain models with multiple interconnected areas, model partitioning and process mapping emerge as critical challenges that directly impact simulation efficiency.
While achieving basic load balancing through uniform distribution of neural units across processes, the widely adopted round-robin partitioning method fundamentally fails to account for the intrinsic communication patterns observed in biological neural networks.
Neuroimaging studies reveal that inter-area connectivity follows a distance-dependent principle where intra-area communication density exceeds inter-area communication density by orders of magnitude.
This biological reality renders the conventional round-robin partition suboptimal, as it leads to a significant amount of avoidable inter-process communication.
To address this mismatch, we implement the area partition method proposed in \cite{zyl2024ATM}, which divides each area into $n$ parts ($n$ varies among areas and is determined by the workload of that area) and designates an exclusive process to simulate each part. 

Process mapping must consider the non-uniform communication costs inherent in modern HPC architectures: the costs of inter-process communication vary significantly based on physical interconnect, with lower costs between processes that are physically closer.
Although processes are logically arranged into a 2D Mesh, their physical deployment follows a linear numbering scheme within computing clusters, resulting in non-uniform communication characteristics between dimensions.
Communication along the X-axis typically occurs between adjacent processes under the same switch hierarchy, exhibiting lower latency.
Conversely, communication along the Y-axis traverses multiple compute nodes across different switch layers, incurring higher communication costs.
Given that communication within brain areas is dense, our mapping strategy prioritizes co-locating processes from the same area along shared Y-axis coordinates.
This design choice underpins the preference for YX routing over XY routing: the latter would introduce additional relay operations for inter-area communication.

\begin{figure}[t]
  \centering
  \begin{subfigure}[b]{0.49\linewidth}
    \includegraphics[width=\textwidth]{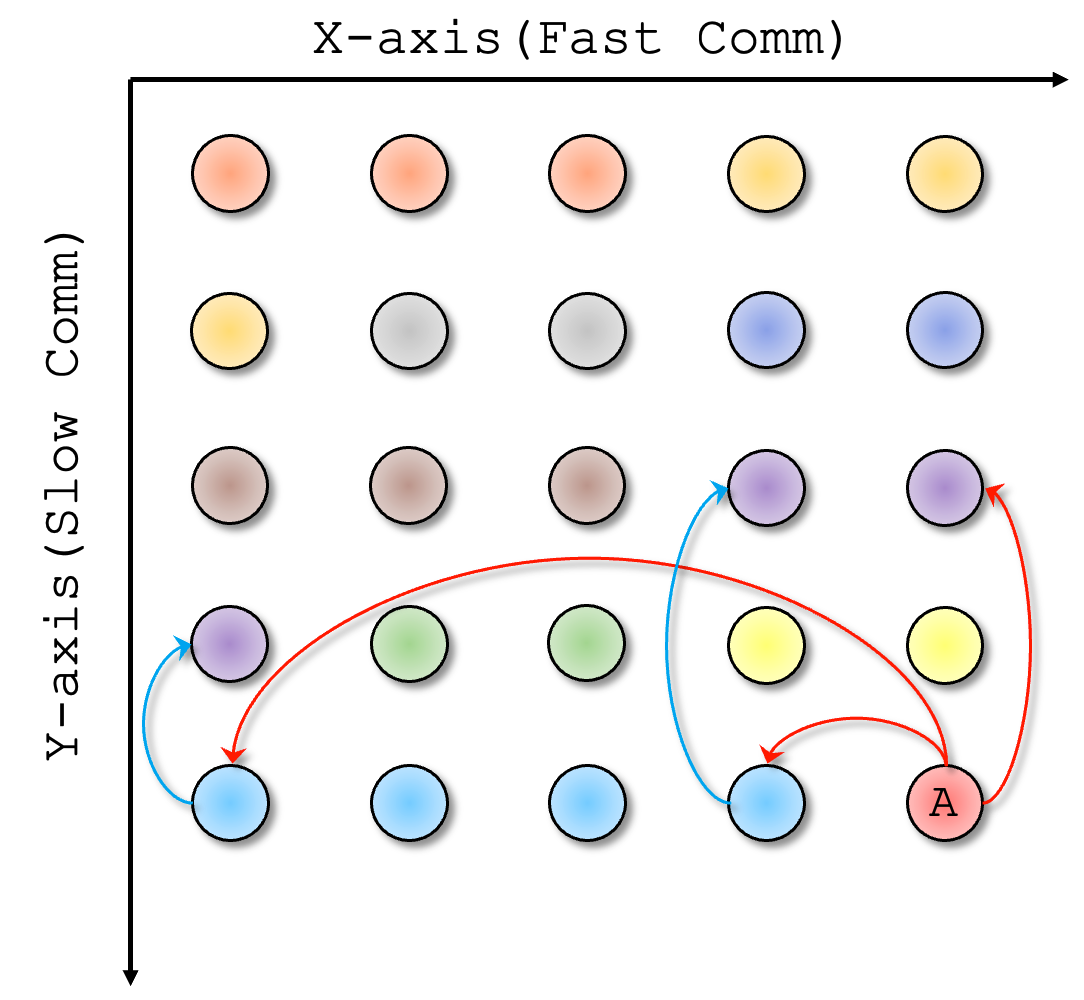}
    \caption{Using XY routing without greedy area-clustering.}
  \end{subfigure}
  \hfill
  \begin{subfigure}[b]{0.49\linewidth}
    \includegraphics[width=\textwidth]{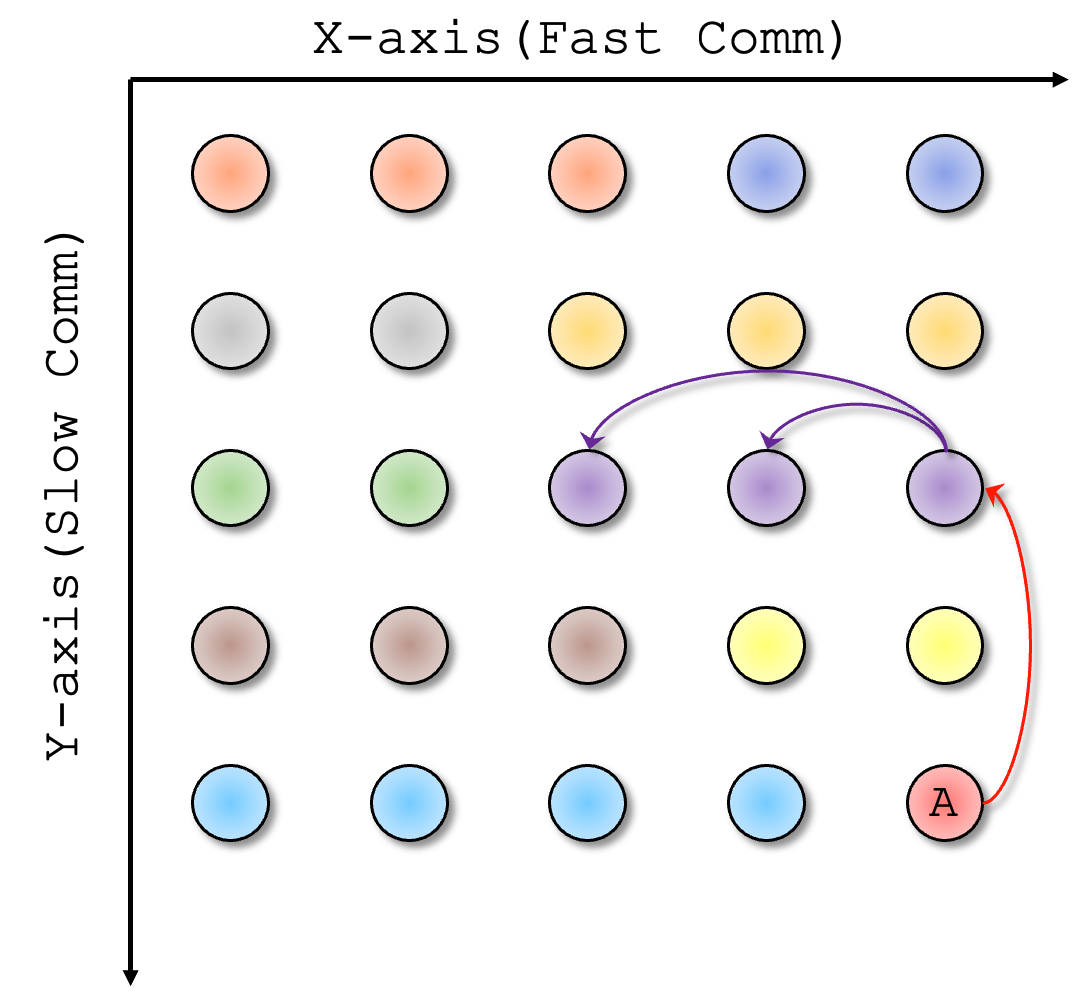}
    \caption{Using YX routing with greedy area-clustering.}
  \end{subfigure}
    \caption{Message routing and process mapping. Each circle represents a process, and different colors represent different areas. Process A of the red area sends spikes to the processes of the purple area. (a) If XY routing without greedy area-clustering is used, sending spikes from A costs 3 slow communications on the Y-axis and 2 fast communications on the X-axis. (b) If YX routing with greedy area-clustering is used, sending spikes from A only costs 1 slow communication on the Y-axis and 2 fast communications on the X-axis.}
    \label{fig:mapping}
\end{figure}

Consequently, we design a greedy area-clustering algorithm that maps communication-intensive parts to nearby processes. Specifically, it prioritizes mapping the same brain area to the same Y-dimension, while different brain areas on the same Y-dimension have denser communication. \Cref{fig:mapping} provides examples of different routing strategies and mapping methods.

\section{Experiments}
\label{sec:experiment}

\subsection{Experiment Setup}

Experiments were conducted on an HPC cluster connected via a 3-layer fat-tree TH Express network. Each compute node is equipped with dual 32-core 2.6 GHz Intel 8358P CPUs, 512 GB memory, and two new 200 Gbps TH Express NICs. Each compute node is configured to host 4 MPI processes. Due to limited cluster resources, we can only use a maximum of 174 compute nodes. The software stack includes GCC 11.4.0, MPICH 4.1.3, and UNR \cite{feng2024unr}. \textbf{OACM} is implemented based on the \textbf{CORTEX} \cite{lyu2024cortex} simulator, which is specially designed for large-scale SNN simulation.

We evaluated performance using two benchmarks: Potjans' cortical microcircuit model \cite{potjans2014cell} and a multi-area marmoset brain model adapted from the CORTEX suite. Specifically, Potjans' model comprises four layers (L2/3 to L6) with $7.7 \times 10^5$ neurons and $2.9 \times 10^{8}$ synapses. The multi-area model spans 116 cortical areas, with $9.7 \times 10^6$ neurons and $3.7 \times 10^{10}$ synapses. Simulations ran for 15,000 timesteps with a resolution of 0.1 ms per step.


\subsection{Correctness Evaluation}

\begin{figure}[t]
  \centering
  \begin{subfigure}[b]{1\linewidth}
    \includegraphics[width=0.48\textwidth]{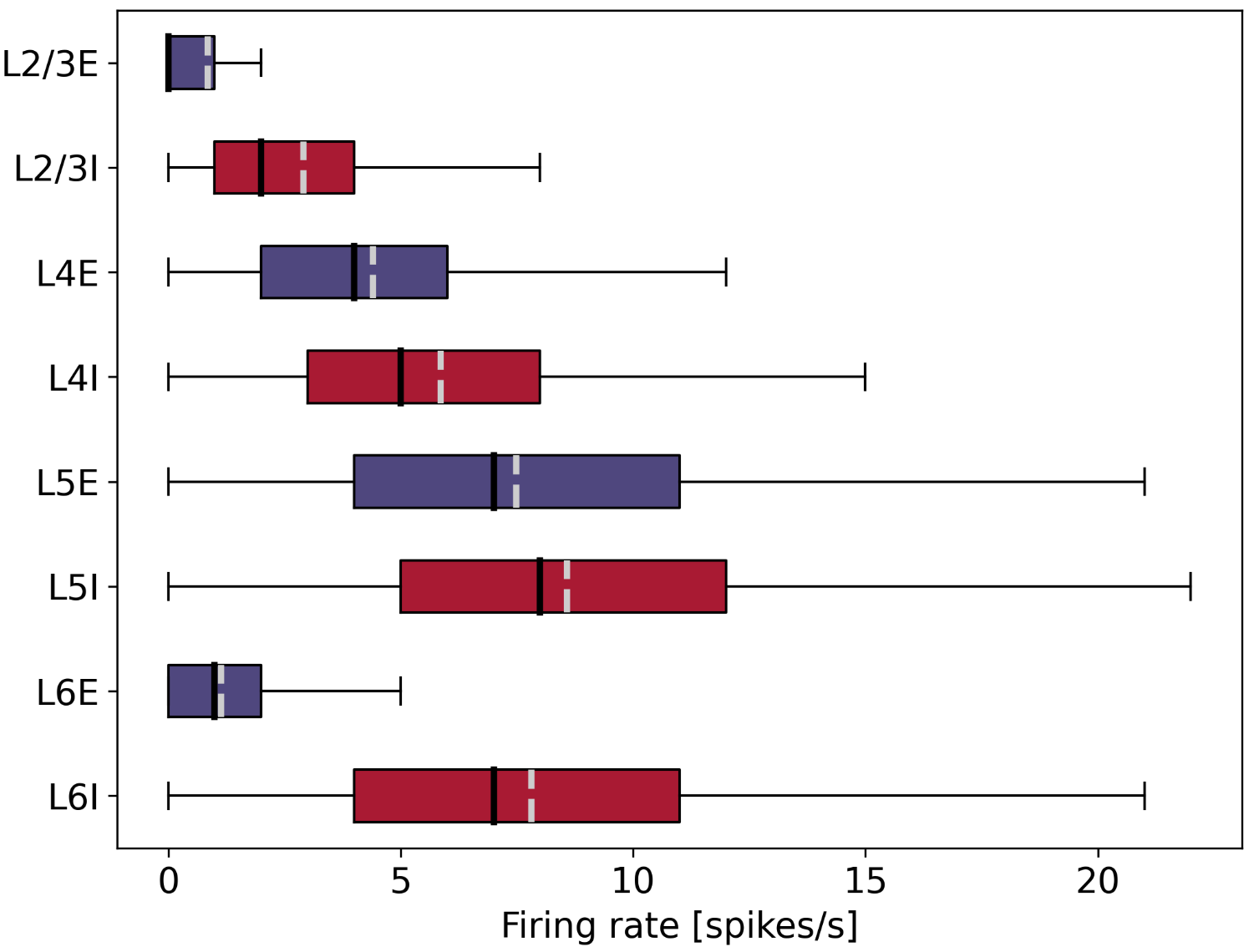}
    \includegraphics[width=0.48\textwidth]{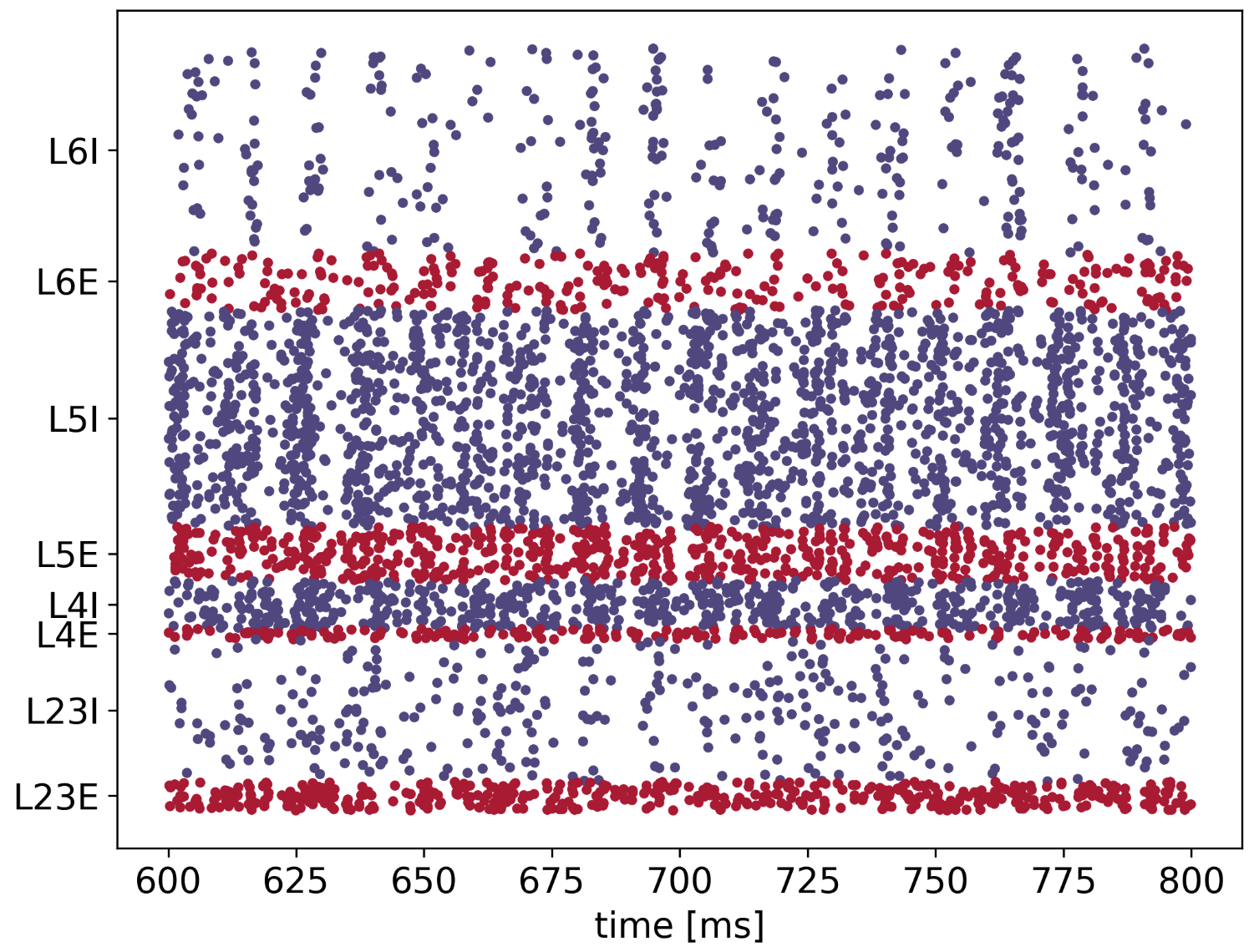}
    \caption{CORTEX's simulation result.}
  \end{subfigure}
  \begin{subfigure}[b]{1\linewidth}
    \vspace{15pt}
    \includegraphics[width=0.48\textwidth]{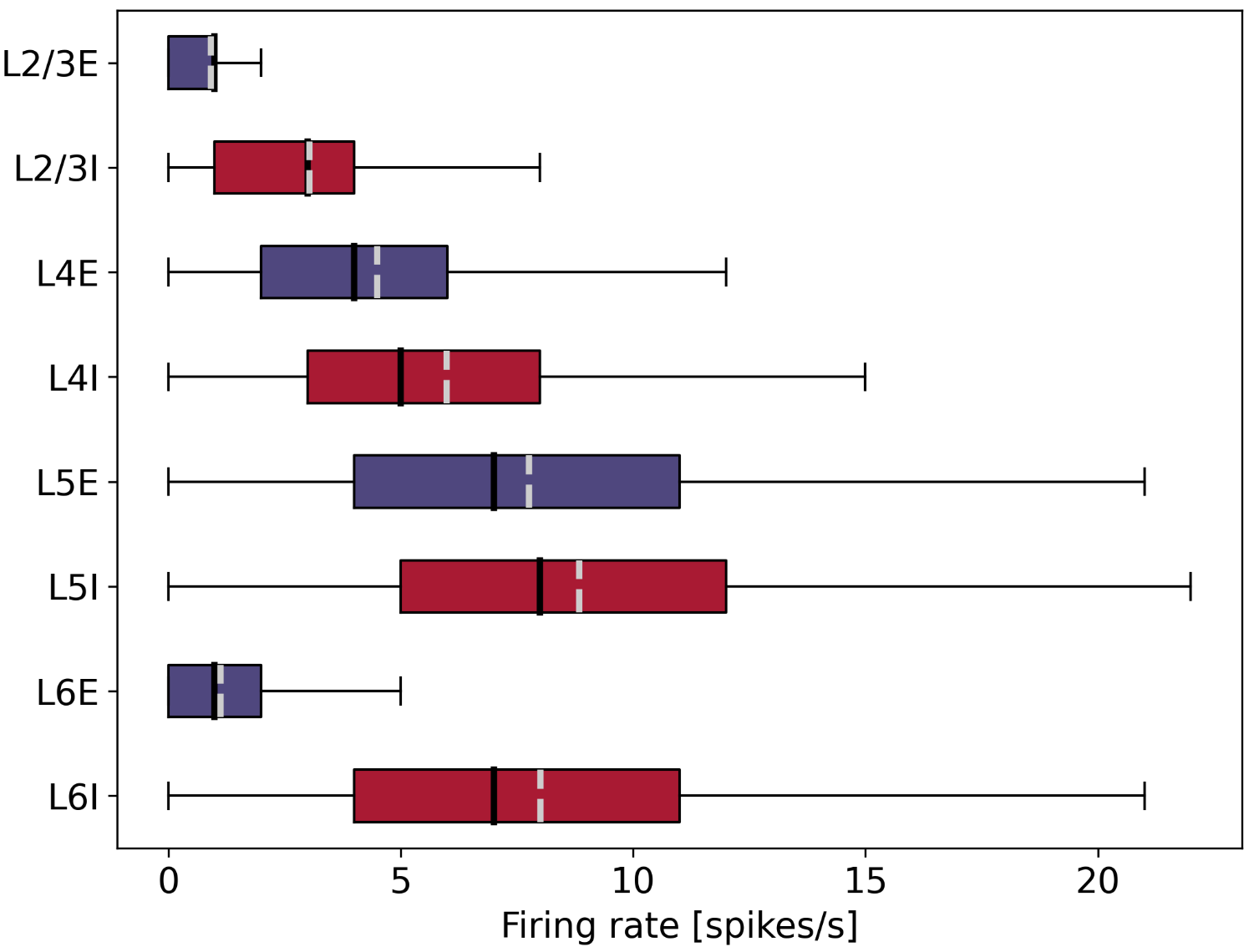}
    \includegraphics[width=0.48\textwidth]{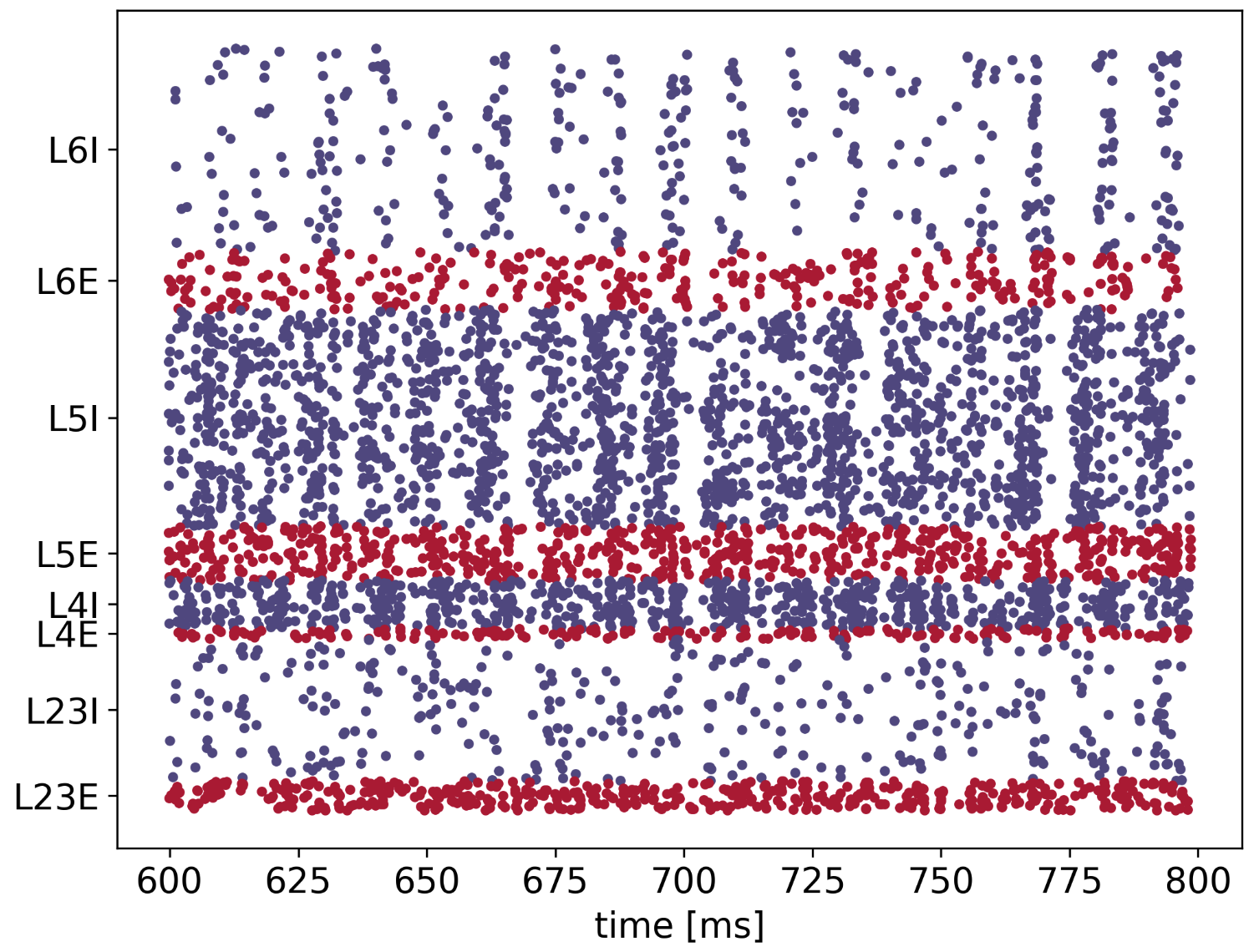}
    \caption{OACM's simulation result.}
  \end{subfigure}
    \caption{The correctness comparison using Potjans' model.}
    \label{fig:correct}
\end{figure}

To validate that our optimization preserves simulation fidelity, we employed Potjans' model following the verification methodology in \cite{Gewaltig:NEST, lyu2024cortex}. We analyzed population-specific firing rates and spike rasters to compare OACM against the baseline CORTEX. As shown in \Cref{fig:correct}, the firing rate distributions (box plots) and spike timings (rasters) of OACM closely match those of CORTEX. Minor discrepancies are observed; however, these are attributable to stochastic factors, such as random connectivity and background noise, and remain within acceptable statistical bounds.

\begin{figure}[t]
  \centering
  \begin{subfigure}[b]{\linewidth}
    \includegraphics[width=\textwidth]{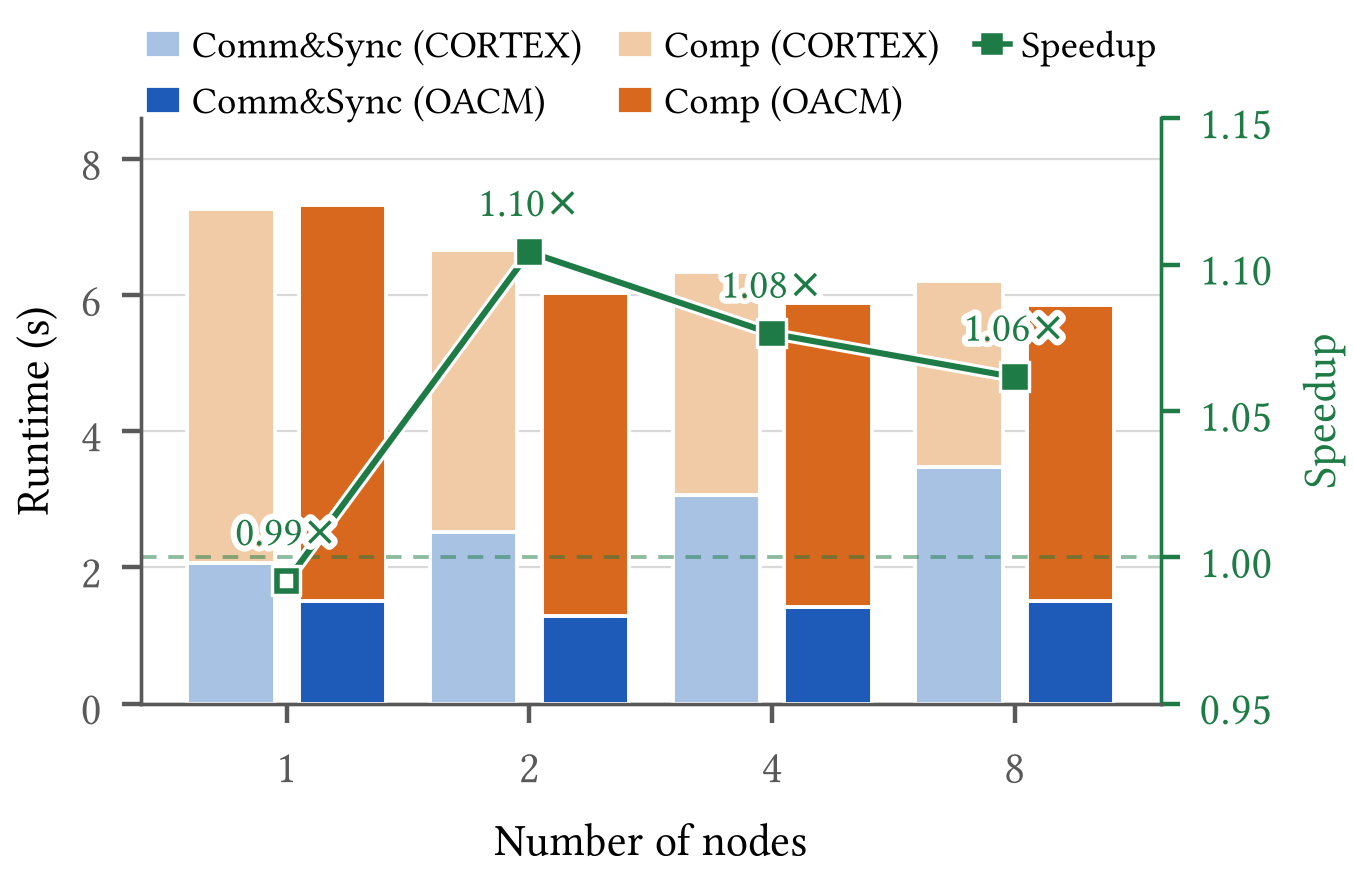}
    \caption{The Potjans' model.}
  \end{subfigure}
  \begin{subfigure}[b]{\linewidth}
    \includegraphics[width=\textwidth]{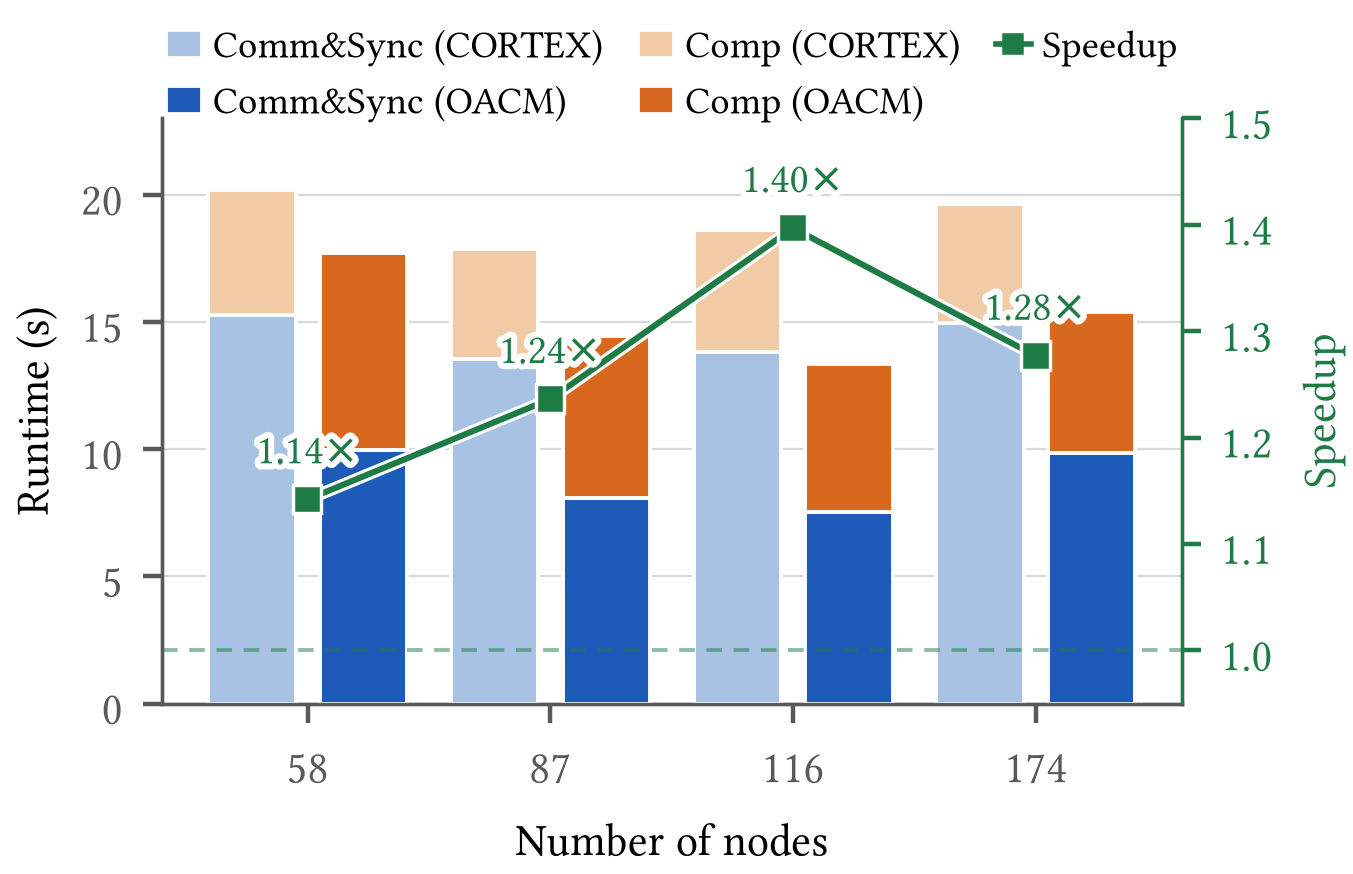}
    \caption{The multi-area model.}
  \end{subfigure}
    \caption{The runtime performance comparison. OACM achieves up to a 1.40$\times$ speedup over CORTEX in the multi-area model. In the small-scale Potjans' model, OACM also attains a 1.10$\times$ speedup at 2 nodes.}
    \label{fig:runtime}
\end{figure}

\begin{figure}[t]
  \centering
  \begin{subfigure}[b]{\linewidth}
    \includegraphics[width=\textwidth]{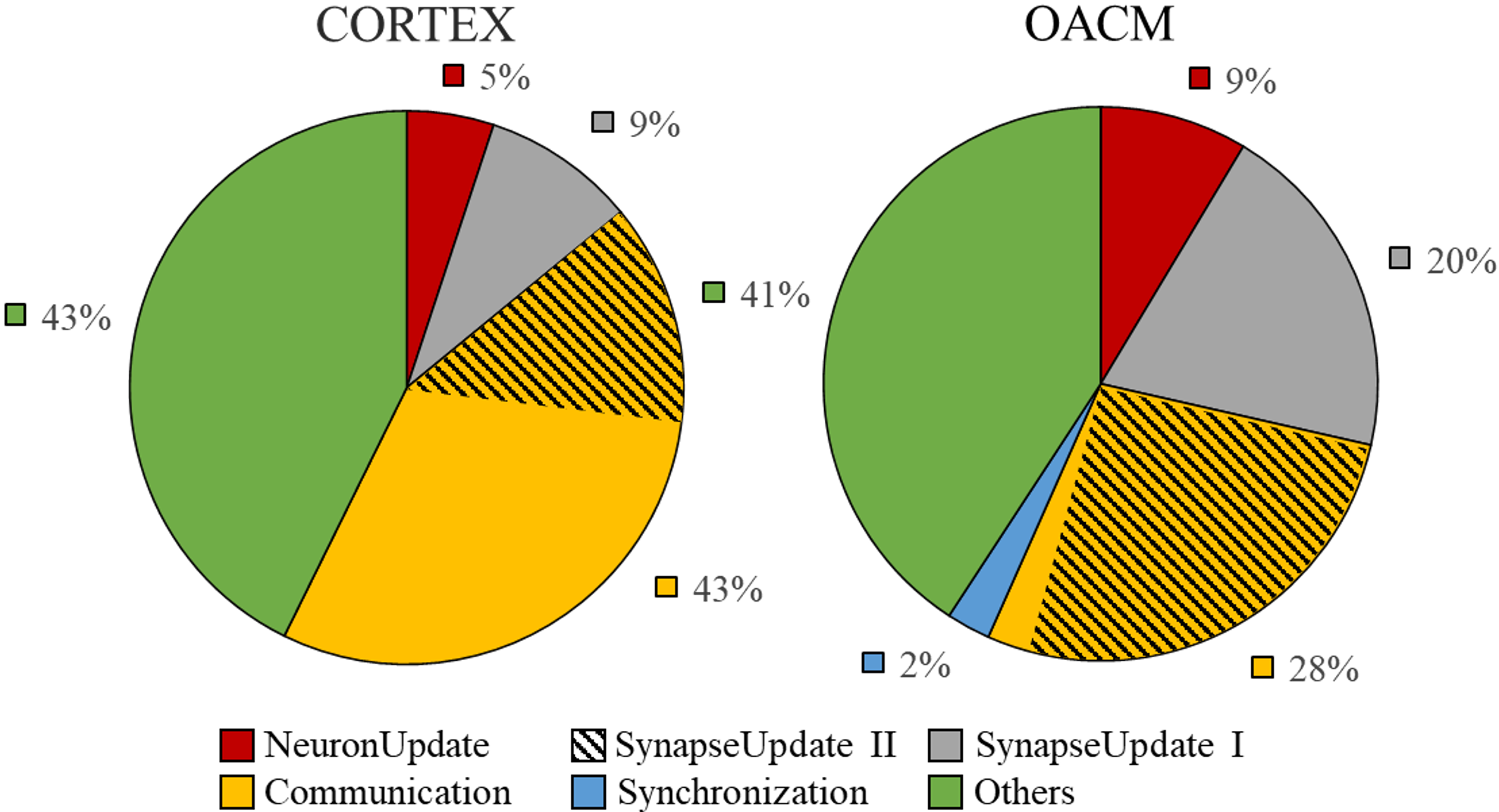}
    \caption{Simulating the Potjans' model with 2 nodes.}
  \end{subfigure}
  \begin{subfigure}[b]{\linewidth}
    \includegraphics[width=\textwidth]{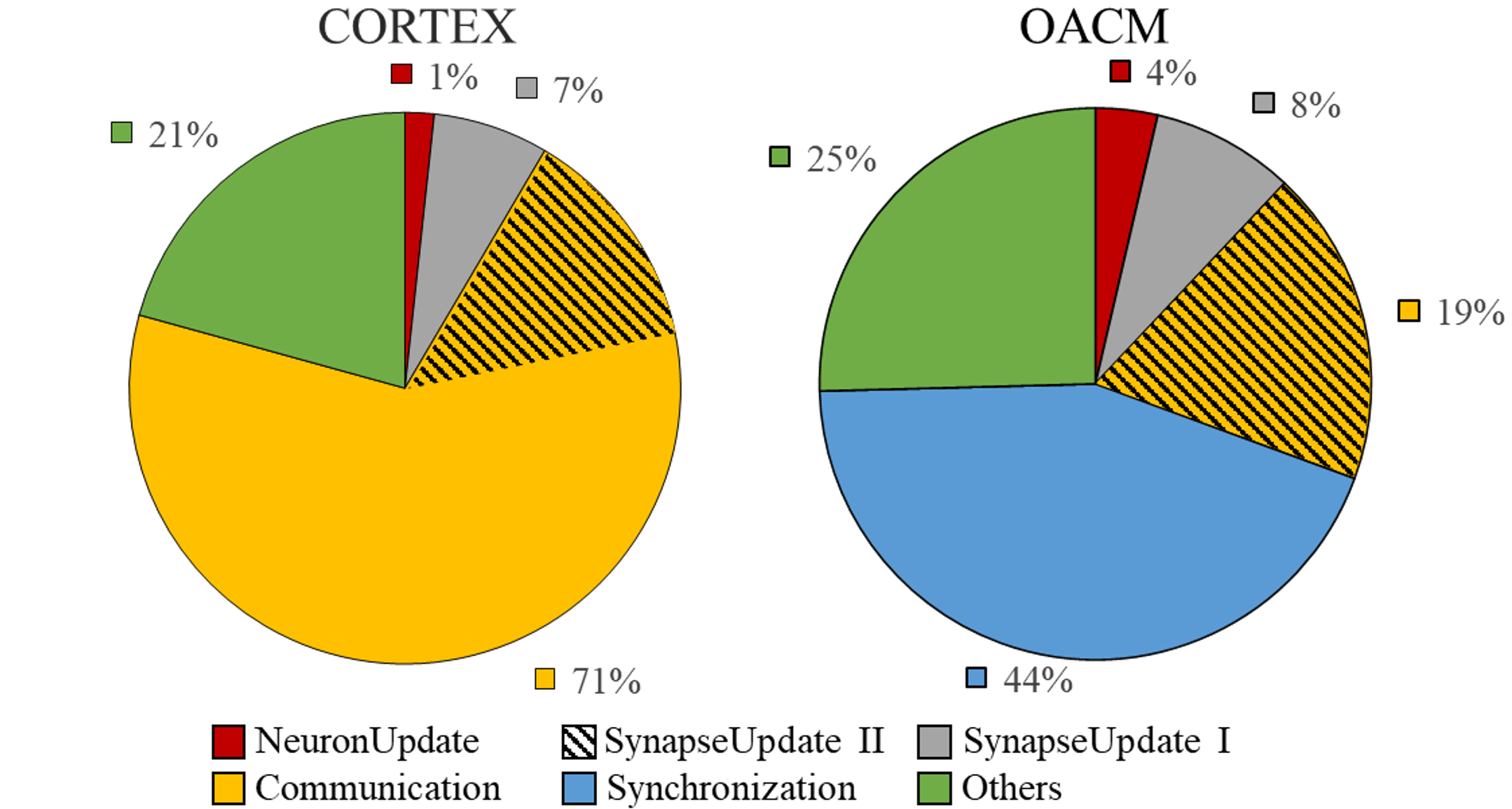}
    \caption{Simulating the multi-area model with 174 nodes.}
  \end{subfigure}
    \caption{The runtime percentage of each function. The black slashed segment, representing the `SynapseUpdate II' used to overlap communication with computation, almost entirely covers the `Communication' portion in OACM, indicating that nearly all of OACM's communication time is hidden by computation. }
    \label{fig:percent}
\end{figure}

\subsection{Performance Evaluation}



The trade-off between synchronization window size and runtime performance is as follows. Results indicate that increasing the window size initially reduces runtime, despite a concurrent rise in the average number of rollbacks. However, extending the window size beyond 20 yields diminishing returns, significantly increasing rollbacks while causing runtime to plateau or slightly increase. This pattern reflects the balance between minimizing synchronization overhead and managing rollback costs. Consequently, a synchronization window size of 20 was selected for our experiments as it provided optimal performance across both models.

Subsequently, we evaluate and compare the performance of CORTEX and OACM. Here, overall runtime is used for comparison without dissecting each optimization method via ablation studies. The reason is that each optimization method is sequentially dependent and only takes effect when combined. \Cref{fig:runtime} illustrates the runtime and relative speedup of OACM and CORTEX. The line chart represents the speedup, while the bar chart depicts the runtime. \Cref{fig:percent} further breaks down the average time consumption of functional components and their proportions across all processes for CORTEX (left) and OACM (right). Since CORTEX employs the MPI\_Allgatherv method for communication, which inherently includes synchronization, it is hard to collect statistics separately. Therefore, we have classified the synchronization part of CORTEX into the communication part. In \Cref{fig:percent}, the black slashed segment over the `Communication' portion represents the percentage of `SynapseUpdate II', which is dedicated to overlapping computation and communication. For OACM, this portion almost entirely covers the asynchronous communication, thereby achieving a complete overlap of computation and communication.

In small-scale simulations of the Potjans' model, the number of nodes varies from 1 to 8. As shown in \Cref{fig:runtime}, there is minimal runtime difference between OACM and CORTEX. OACM underperforms slightly at 1 node but achieves a 1.1$\times$ speedup at 2 nodes. As shown in \Cref{fig:percent}(a), communication and synchronization are not dominant bottlenecks at smaller node counts. Additionally, OACM introduces extra computational overhead, such as rollback operations and anti-spike processing. Given its communication-intensive nature and all-to-all inter-process communication pattern, the model aligns well with CORTEX's MPI\_Allgatherv primitive. As the number of nodes increases, the 2D Mesh-based routing mechanism fails to deliver messages in a timely manner, leading to more rollback operations and subsequent performance degradation.

In contrast, the multi-area model exhibits dense intra-area communication but sparse inter-area communication, with irregular connections between areas causing irregular inter-process communication among many processes. In this regime, OACM's advantages become pronounced. When simulating the multi-area model, the number of nodes scales from 58 to 174. CORTEX's simulation time initially decreases with more nodes but increases at 116 nodes due to synchronous communication bottlenecks. By contrast, OACM shows better scalability and outperforms CORTEX at various node counts. The speedup ratio rises with the number of nodes but declines at 174 nodes, peaking at 1.4$\times$ at 116 nodes. As shown in \Cref{fig:percent}(b), communication and synchronization account for over 70\% of CORTEX's runtime, whereas OACM significantly reduces communication and synchronization time despite introducing additional computational overhead. Synchronization overhead reduction is the main driver of OACM's performance gain over CORTEX.

Furthermore, we compare the runtime performance between OACM and NEST \cite{Gewaltig:NEST} using the multi-area model. NEST is an open-source, MPI-enabled simulator specialized for large-scale SNN simulations. It is widely adopted in computational neuroscience and brain-inspired computing research, including the Human Brain Project. To present the data in a visually appealing manner, we give this comparison in tabular form, with the results shown in \Cref{tab:nest}. When the number of nodes increased from 58 to 174, NEST exhibited poor scalability and runtime performance, with runtime increasing significantly. In contrast, OACM demonstrates superior scalability and runtime performance, achieving speedups of more than 4$\times$ across different node counts. Specifically, with 174 nodes, OACM achieves a speedup of 22.6$\times$ compared to NEST. This indicates that OACM is significantly more efficient and scalable than NEST, making it a more suitable choice for large-scale SNN simulations.

\begin{table}[t]
  \caption{Runtime performance comparison with NEST}
  \label{tab:nest}
  \centering
  \begin{tabular}{|c|c|c|c|}
    \hline
    Number of Nodes & NEST runtime & OACM runtime & Speedup\\
    \hline
    58 & 72.08s & 17.99s & 4$\times$ \\
    \hline
    87 & 89.39s & 15.84s & 5.6$\times$\\
    \hline
    116 & 124.62s & 13.83s & 9$\times$\\
    \hline
    174 & 295.1s & 13.01s & 22.6$\times$\\
    \hline
\end{tabular}
\end{table}

\section{Conclusion}
\label{sec:conclusion}


In this paper, we present OACM, an Optimistic Asynchronous Communication Model that addresses the fundamental synchronization and communication bottlenecks in large-scale SNN simulations.
By leveraging the biological characteristics of SNNs -- sparse spiking activity and bounded synaptic delays -- we strategically integrate Optimistic PDES techniques into time-driven simulation frameworks.
This hybrid approach reduces synchronization frequency while maintaining correctness through localized rollback mechanisms within synchronization windows, preserving the simplicity of implementation and the parallelization advantages of time-driven methods.

The introduced rollback mechanism's sensitivity to message transmission delays necessitated advanced communication optimizations.
We employ asynchronous one-sided communication via the UNR library to minimize communication latency and achieve complete computation-communication overlap, while managing receive buffers through application-level flow control.
Furthermore, to address the bandwidth underutilization caused by numerous small messages, we implement adaptive message aggregation and routing within a 2D-HyperX virtual topology, consolidating small packets into larger transmissions and delivering messages through at most one forwarding hop.

Experimental evaluation demonstrates that OACM achieves significant performance improvements, with up to 1.4$\times$ speedup over the original CORTEX simulator and more than 22.6$\times$ speedup compared to NEST when simulating the multi-area marmoset brain model at scales of up to 174 compute nodes.
These results validate our approach's effectiveness in overcoming the synchronization and communication bottlenecks that have long constrained large-scale SNN simulations.

\section{Future Work}
\label{sec:future}


While our work successfully integrates time-driven and event-driven approaches to optimize communication efficiency, the current rollback mechanism does not yet explicitly support evolving synaptic states, such as those introduced by synaptic plasticity. Extending OACM to richer synaptic dynamics introduces additional temporal dependencies, since synaptic states may evolve in response to spike history and must remain causally consistent with neuronal states. Future work will therefore investigate lightweight mechanisms for jointly checkpointing, rolling back, and restoring neuronal and synaptic states within synchronization windows, while controlling the associated memory and recomputation overhead. This extension would enable OACM to support more biologically sophisticated large-scale SNN simulations.

\printcredits

\section*{Declaration of interests}
The authors declare that they have no known competing financial interests or personal relationships that could have appeared to influence the work reported in this paper.

\bibliographystyle{elsarticle-num}

\bibliography{reference/Reference}



\end{document}